\documentclass[fleqn,usenatbib]{mnras}
\usepackage{fancyhdr}  
\usepackage{newtxtext,newtxmath}
\usepackage[T1]{fontenc}
\DeclareRobustCommand{\VAN}[3]{#2}
\let\VANthebibliography\thebibliography
\def\thebibliography{\DeclareRobustCommand{\VAN}[3]{##3}\VANthebibliography}
\usepackage{graphicx}
\usepackage{amsmath}
\usepackage{hyperref} 
\usepackage{orcidlink}

\title[LMC and SMC in diagnostic diagrams]{SDSS-V LVM: Spatial Resolution Effects on Diagnostic Diagrams in the Magellanic Clouds}

\author[A. Z. Lugo-Aranda et al.]{
A. Z. Lugo-Aranda\orcidlink{0000-0001-9226-9178}$^{1}$\thanks{E-mail: alugo@astro.unam.mx},
Lesly C. Castañeda\orcidlink{0009-0000-2341-9865}$^{2}$,
H. J. Ibarra-Medel\orcidlink{0000-0002-9790-6313}$^{2}$,
C. Román-Zúñiga\orcidlink{0000-0001-8600-4798}$^{1}$,\newauthor
J. E. Méndez-Delgado\orcidlink{0000-0002-6972-6411}$^{2}$,
Sebastián F. Sánchez\orcidlink{0000-0001-6444-9307}$^{1,3}$,
Rodolfo de J. Zermeño\orcidlink{0009-0009-0081-4323}$^{2}$,
Mónica A. Villa-Durango\orcidlink{0009-0008-1605-4771}$^{1}$,\newauthor
Irene Cruz-González\orcidlink{0000-0002-2653-1120}$^{2}$,
Aida Wofford\orcidlink{0000-0001-8289-3428}$^{1}$,
Oskar Aranguré\orcidlink{0009-0004-7070-1782}$^{1, 4}$,
G. A. Blanc\orcidlink{0000-0003-4218-3944}$^{5, 6}$,
Priyanka Chakraborty\orcidlink{0000-0002-4469-2518}$^{7}$,\newauthor
Evgeniya Egorova\orcidlink{0000-0003-2717-8784}$^{8}$,
Oleg V. Egorov\orcidlink{0000-0002-4755-118X}$^{8}$,
Thomas Hilder\orcidlink{0000-0001-7641-5235}$^{9}$,
Evelyn J. Johnston\orcidlink{0000-0002-2368-6469}$^{10}$,
Amy M. Jones\orcidlink{0000-0002-2262-8240}$^{4}$,\newauthor
P. Jiménez-Hernández\orcidlink{0000-0002-6779-5181}$^{1}$,
K. Kreckel\orcidlink{0000-0001-6551-3091}$^{8}$,
Alfredo J. Mej\'{i}a-Narv\'{a}ez\orcidlink{0000-0002-8931-2398}$^{11}$,
Christophe Morisset\orcidlink{0000-0001-5801-6724}$^{12}$,\newauthor
Rogelio Orozco-Duarte\orcidlink{0000-0003-2193-3005}$^{13}$,
Alexandre Roman-Lopes\orcidlink{0000-0002-1379-4204}$^{14}$,
José G. Fernández-Trincado\orcidlink{0000-0003-3526-5052}$^{15}$
\\
$^{1}$Instituto de Astronomía, Universidad Nacional Autónoma de México, A.P. 106, Ensenada 22800, BC, México\\
$^{2}$Instituto de Astronomía, Universidad Nacional Autónoma de México, A.P. 70-264, 04510 México D. F., México\\
$^{3}$Instituto de Astrofísica de Canarias, Vía Láctea s/n, 38205 La Laguna, Tenerife, Spain\\
$^{4}$Space Telescope Science Institute, 3700 San Martin Drive, Baltimore, MD, 21218, USA\\
$^{5}$Observatories of the Carnegie Institution for Science, 813 Santa Barbara Street, Pasadena, CA 91101, USA\\
$^{6}$Departamento de Astronom\'{i}a, Universidad de Chile, Camino del Observatorio 1515, Las Condes, Santiago, Chile\\
$^{7}$Center for Astrophysics | Harvard \& Smithsonian, 60 Garden St., Cambridge, MA 02138, USA\\
$^{8}$Astronomisches Rechen-Institut, Zentrum f\"{u}r Astronomie der Universit\"{a}t Heidelberg, M\"{o}nchhofstra\ss e 12-14, D-69120 Heidelberg, Germany\\
$^{9}$School of Physics and Astronomy, Monash University VIC 3800, Australia\\
$^{10}$Universidad Diego Portales, Instituto de Estudios Astrofísicos, Facultad de Ingeniería y Ciencias, Av. Ejército Libertador 441, Santiago, Chile\\
$^{11}$Departamento de Astronom\'ia, Universidad de Chile, Av. Libertador Bernardo O'Higgins 1058, Santiago de Chile\\
$^{12}$Universidad Nacional Autónoma de México, Instituto de Ciencias F\'isicas, Av. Universidad S/N, Chamilpa, 62200 Cuernavaca, Morelos, México\\
$^{13}$Instituto de Radioastronomía y Astrofísica, Universidad Nacional Autónoma de México, 58089 Morelia, Michoacán, Mexico\\
$^{14}$Department of Astronomy, Universidad de La Serena, Av. Raul Bitran \#1302, La Serena, Chile\\
$^{15}$Centro de investigación en Astronomía, Facultad de Ingeniería, Ciencia y Tecnología, Universidad Bernardo O’Higgins, Av. Viel 1497, Santiago, 8370993, Chile\\
}
\date{Accepted XXX. Received YYY; in original form ZZZ}

\pubyear{2026}

\begin{document}
\label{firstpage}
\pagerange{\pageref{firstpage}--\pageref{lastpage}}
\maketitle

\begin{abstract}
We investigate how spatial resolution affects emission-line diagnostics in the interstellar medium (ISM) of the Small and Large Magellanic Clouds (SMC and LMC) using SDSS Local Volume Mapper (LVM) data obtained at and degraded to $\sim$19, $\sim$42, $\sim$170 pc (for the SMC) and $\sim$17, $\sim$68, $\sim$272 pc (for the LMC), respectively. We compare the distributions of strong emission-line ratios through 1D and 2D Kolmogorov–Smirnov tests. We deduced that diagnostics involving ionization and density sensitive ratios such as [OIII]$\lambda$5007/H$\beta$, [SII]$\lambda$$\lambda$6717+6731/H$\alpha$ and [NII]$\lambda$6584/H$\alpha$ are the most strongly impacted by resolution loss in this order. The SMC shows significant deviations even from intermediate scales ($\sim$42 pc), whereas for the LMC, the coarsest resolution ($\sim$272 pc) remains relatively stable, indicating that the SMC is intrinsically more affected by spatial mixing. Resolution degradation produces a marked homogenization effect: extreme values present at the finest scales are progressively diluted, causing diagnostic diagrams to converge toward their mean loci. Diffuse ionized gas is particularly sensitive to this mixing, while H$\alpha$-selected ionized regions exhibit more robust behavior across resolutions. The sensitivity of diagnostic diagrams involving emission-line ratios, equivalent widths, and kinematic properties varies with both spatial resolution and galaxy, with the SMC generally showing stronger changes than the LMC. Our results demonstrate that sub-50 pc resolution is required to resolve the detailed structure of low-metallicity galaxies, and they highlight at least the need for careful interpretation of line-ratio diagnostics in extragalactic studies where coarse spatial resolution is unavoidable.
\end{abstract}


\section{Introduction}

Ionized gas plays a central role in understanding galaxy evolution, providing crucial insights into feedback processes and the physical conditions of the interstellar medium (ISM). Over the past decade, there has been increasing reliance on strong optical emission lines, both recombination lines (RLs) such as H$\alpha$ and H$\beta$, and collisionally excited lines (CELs) such as [OIII]$\lambda$5007, [SII]$\lambda$$\lambda$6717, 6731, [NII]$\lambda$6584, and [OI]$\lambda$6300 as diagnostic tools for probing and identifying ionization conditions in extragalactic HII regions \citep{bergskillmancroxall2015, sanchezgalbanyperez2015, shapleysanderstopping2025}. A major challenge arising from the rapid improvement in spatial resolution is the proper interpretation of diagnostics in a spatially resolved manner, as these were originally developed and calibrated using data with much coarser resolution \citep{kewleydopitasutherland2001, kauffmanheckmantremonti2003, cidfernandesstasinskamateus2011, sanchezlugosanchezalmeida2024}. This issue is particularly relevant when attempting to differentiate between the various ionization sources of the ISM \citep{espinosasanchezmorisset2020}.

\begin{figure*} 
\centering
\includegraphics[width=1\textwidth]{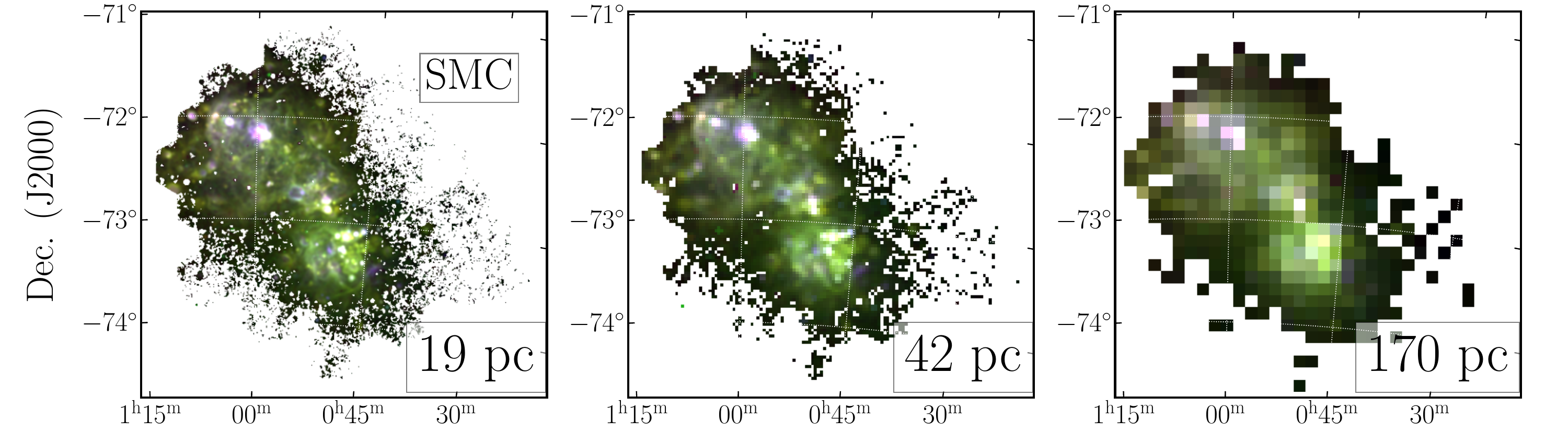}
\includegraphics[width=1\textwidth]{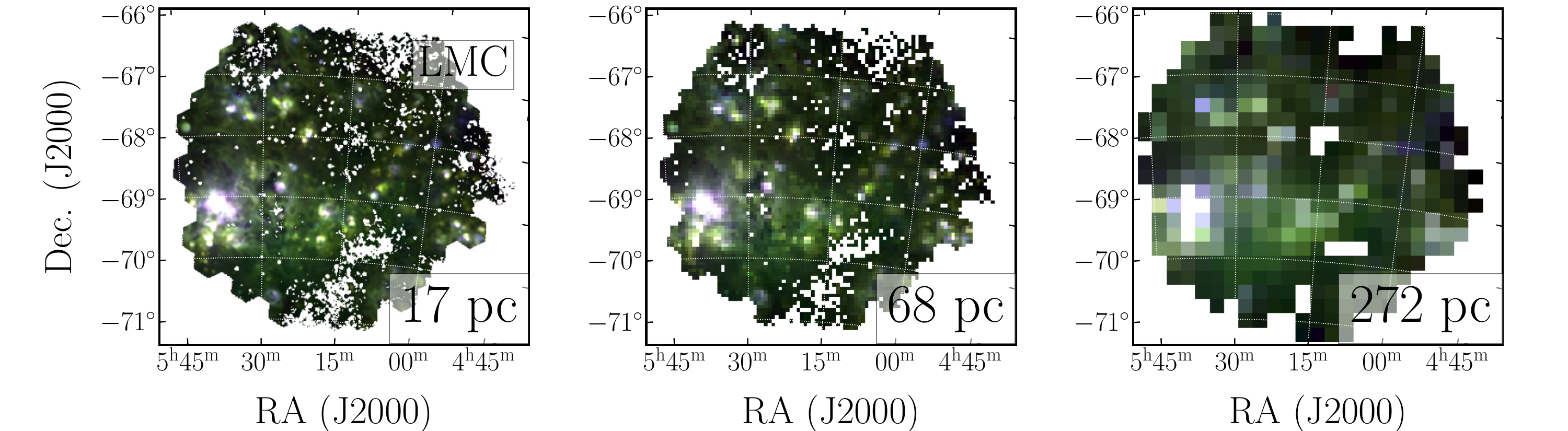}
\caption{RGB images of the Magellanic Clouds.
The Small Magellanic Cloud (upper panels) is shown from left to right at spatial scales of $\sim$19, $\sim$42, and $\sim$170 pc, while the Large Magellanic Cloud (lower panels) is shown from left to right at $\sim$17, $\sim$68, and $\sim$272 pc. In each RGB image, the red, green, and blue channels correspond to the flux intensities of [NII]$\lambda$6584, H$\alpha$, and [OIII]$\lambda$5007, respectively.} 
\label{fig:MC_RGB}
\end{figure*}

In the wavelength range covered by the optical regime, the major advances have been driven largely by modern integral-field spectroscopic surveys or programs. These include observations of nearby galaxies with the Multi Unit Spectroscopic Explorer (MUSE), such as those from the All-weather MUse Supernova Integral-field of Nearby Galaxies survey (AMUSING++; \citealp{lopezsanchezanderson2020}), the Physics at High Angular Resolution in Nearby Galaxies program (PHANGS-MUSE; \citealp{emsellemschinnerersantoro2022}), mosaicking of large targets (\citealp{feltrebelfiorecresci2026}), the Measuring Active Galactic Nuclei Under MUSE Microscope survey (MAGNUM; \citealp{magnum1, magnum2}), and, more recently, by the Local Volume Mapper (LVM; \citealp{droryblanckreckel2024}). The LVM, a key program of the Sloan Digital Sky Survey (SDSS-V; \citealp{kollmeierrixaerts2025}), maps approximately 4300 square degrees of the sky, covering the southern Milky Way, the Magellanic Clouds (MCs), and other Local Group systems. LVM is a wide-field integral-field spectroscopic survey providing optical spectroscopy over a wavelength range of 3600–9800 \AA, with 35.3" diameter science fibers and an IFU field of view of approximately 0.5' in diameter. The high spatial resolution provided by the LVM (on the order of 0.1 pc for Milky Way objects and on the order of tens of parsecs for the Magellanic Clouds) reveals small-scale structures and allows us to distinguish ionization sources that would appear mixed or blended at degraded resolutions \citep{droryblanckreckel2024}. Its capabilities enable detailed studies of star-forming regions (e.g. \citealp{kreckelegorovegorova2024, villadurangobarreraroman2025, sattlermendezkreckel2025, hildercaseydalcaton2025, singhblanckumari2026}), supernova remnants \citep{sarbadhicarylongraymond2025}, Wolf–Rayet stars \citep{gonzalestorasanderegorova2025}, the planetary nebulae \citep{sanchezmendezmejia2026}, among other topics.

High spatial resolution observations also offer critical benchmarks for interpreting the unresolved emission of distant galaxies \citep{carnallmcluredunlop2023, carnianihainlinedeugenio2024}, including those at high redshift revealed by the James Webb Space Telescope \citep{arellanobergchisholm2022, arellanocullencarnall2025, rogersstromrudie2024, rogersstromrudie2025, stantoncullencarnall2025}. By studying the ionized ISM in a spatially resolved manner in the local Universe, we can better understand the physical processes that influence the diagnostic tools applied to extragalactic systems where spatial resolution is inherently limited. Understanding the capabilities and limitations of emission-line diagnostics in nearby, well-resolved environments is therefore essential for bridging the gap between local and distant galaxy studies.

Among extragalactic systems, those with low metallicities, such as the Magellanic Clouds with metallicities of Z $<$ 0.5 Z${\odot}$ for the LMC and Z $<$ 0.2 Z${\odot}$ for the SMC; \citealp{russelldopita1992}), are particularly important for understanding the processes that drive galaxy evolution, since their chemical composition resembles that of galaxies at early cosmic epochs \citep{izotovthuan2024}. The integral-field spectroscopy provided by the LVM enables detailed studies of their properties, particularly how emission-line diagnostics change when the gas exhibits lower cooling efficiencies, a harder ionizing radiation field, and complex interactions with the underlying stellar populations. These galaxies, situated at distances of approximately 50–60 kpc \citep{scowcriftfreedmanmadore2016, pietrzynskigraczykgallenne2019}, are the most massive dwarf satellite galaxies of the Milky Way. Owing to their proximity and respective spatial orientations or inclinations (estimated using red clump stars), with i$_{\rm LMC} \approx 23.4^{\circ}$ (e.g., \citealp{choiolsenbesla2022}) and i$_{\rm SMC} \approx 35.4^{\circ}$ (e.g., \citealp{tattonvanlooncioni2021}), they offer the unique advantage of resolving individual stars \citep{gochermannschmidt2002, mucciarelliminellibellazzini2023} as well as key structures of the interstellar medium, such as molecular clouds \citep{paradismenyjuvela2019, saldanorubiobolatto2024} and regions shaped by stellar feedback \citep{mccluredenesdickey2018, grishuninweisscolombo2024}, including HII regions \citep{lopezkrumholzbolatto2014} and supernova remnants \citep{toribiodominguezesteban2017, maggifilipovicvukotic2019}.

The MCs exhibit a broad range of ISM conditions across multiple phases. Their environments span from cold molecular gas reservoirs (e.g.,\citealp{galametzschrubadebruck2020}) to cold and warm dust components, the latter revealed by large spatial detail in recent JWST observations of regions such as 30 Doradus \citep{rodriguezlopezlancaster2025}. They also host diverse radiation fields traced by the large population of ionized nebulae and massive star-forming regions \citep{gallianohonybernard2011, whelanlebougalliano2013}. In addition, the MCs contain extensive hot, X-ray emitting gas produced by stellar associations and supernovae remnants, highlighting the strong impact of stellar feedback on their ISM structure and energetics \citep{kavanaghsasakipoints2012, townsleybroospovich2024}.

Although the MCs likely entered the Milky Way’s virial radius only 1–2 Gyr ago (an event associated with the formation of the Magellanic Stream, \citealp{lucchinidonghiafox2021}), they have long been engaged in mutual gravitational interactions \citep{cullianemackeydacosta2023}. These long-lasting interactions have significantly shaped both their stellar components and their ionized gas, reinforcing their role as local analogs for exploring galaxy evolution in low-mass, metal-poor environments.

Given this diversity in interstellar conditions and the influence of prolonged dynamical interactions, identifying the dominant ionization sources across different regions of the MCs is a key step toward understanding their star formation processes, the role of feedback, and the interplay between stars, gas, and gravitational forces. Various parameters can be employed, including morphology, gas and stellar kinematics, chemical abundances, metallicity distributions, and emission-line ratios. 

In this work, we focus on six different emission-line ratio diagnostics to investigate how different ionization mechanisms contribute across the MCs at multiple spatial resolutions. These diagrams are widely used and provide valuable, though not exclusive, tools for disentangling the contributions of photoionization by young stars, shocks induced by interactions or feedback, and emission from evolved stellar populations \citep{sanchez2020}. Since diagnostic diagrams are applied over diverse spatial scales, we investigate them at the highest available LVM resolution as well as at degraded resolutions. This allows us to quantify the effect of spatial resolution on interpreting ionization sources in extragalactic systems. While powerful for identifying dominant ionization sources, diagnostic diagrams must be complemented by additional physical parameters to obtain a more comprehensive picture, e.g., of the physical conditions of the ionized gas and its stellar populations. This effort is being pursued through complementary approaches, including stellar population analysis by Zermeño et al. (in prep.) and the study of the physical properties of the ionized gas (e.g., temperature and density estimates) by Corina-Castañeda et al. (in prep.).

The structure of this article is as follows: Section 2 describes the observations; Section 3 presents the methodologies and tools used in the analysis; Section 4 discusses the physical interpretation and significance of the different diagnostic diagrams; Section 5 applies these diagrams to various spatial resolutions and cases (all pixels, H$\alpha$-bright regions, and diffuse outer regions) and provides a quantitative characterization of the differences between distributions; and, finally, Sections 6 and 7 present the discussion and conclusions, respectively.

In this work, we adopt a standard $\Lambda$CDM cosmology characterized by the following parameters: $H_0 = 71$ km s$^{-1}$ Mpc$^{-1}$, $\Omega_M = 0.27$, and $\Omega_\Lambda = 0.73$.

\section{LVM observations}

The LVM uses four telescopes located at Las Campanas Observatory (LCO): one dedicated to the calibrated flux system, two dedicated to sky modeling and background subtraction, and one dedicated to the fiber system for scientific observations. The science telescope has a diameter of 16.1 cm and is equipped with a wide-field integral field unit (IFU) containing 1801 microlens-coupled science fibers arranged into 25 hexagonal rings. As each fiber has a 35.3\arcsec\ diameter, and the field has an 88\% filling factor, the total IFU bundle covers 0.5\arcmin\ in diameter. The observations of the Magellanic System are designed to achieve an emission-line sensitivity of 5$\sigma$ = 2 $\times$ 10$^{-18}$ erg s$^{-1}$ cm$^{-2}$ arcsec$^{-2}$ at 6563 \AA\ \citep{herbstbizenbergblanc2024}.

The IFU output is divided into three sections, each feeding a separate spectrograph designed similarly to those of the Dark Energy Spectroscopic Instrument (DESI; \citealp{perruchotguyleguillou2018}).
The spectrographs cover a wide spectral range from 3600 to 9800\AA~with a spectral resolution of $R \approx 4000$ at H$\alpha$, and are split into three dichroic channels: (i) a blue channel (3600–5800\AA), (ii) a red channel (5750–7570\AA), and (iii) an infrared channel (7520–9800\AA), each using Volume Phase Holographic (VPH) gratings \citep{droryblanckreckel2024}. For a full description of the LVM, see \cite{konidarisdroryfroning2020}.
  
\begin{figure*}
    \centering \includegraphics[width=0.99\textwidth]{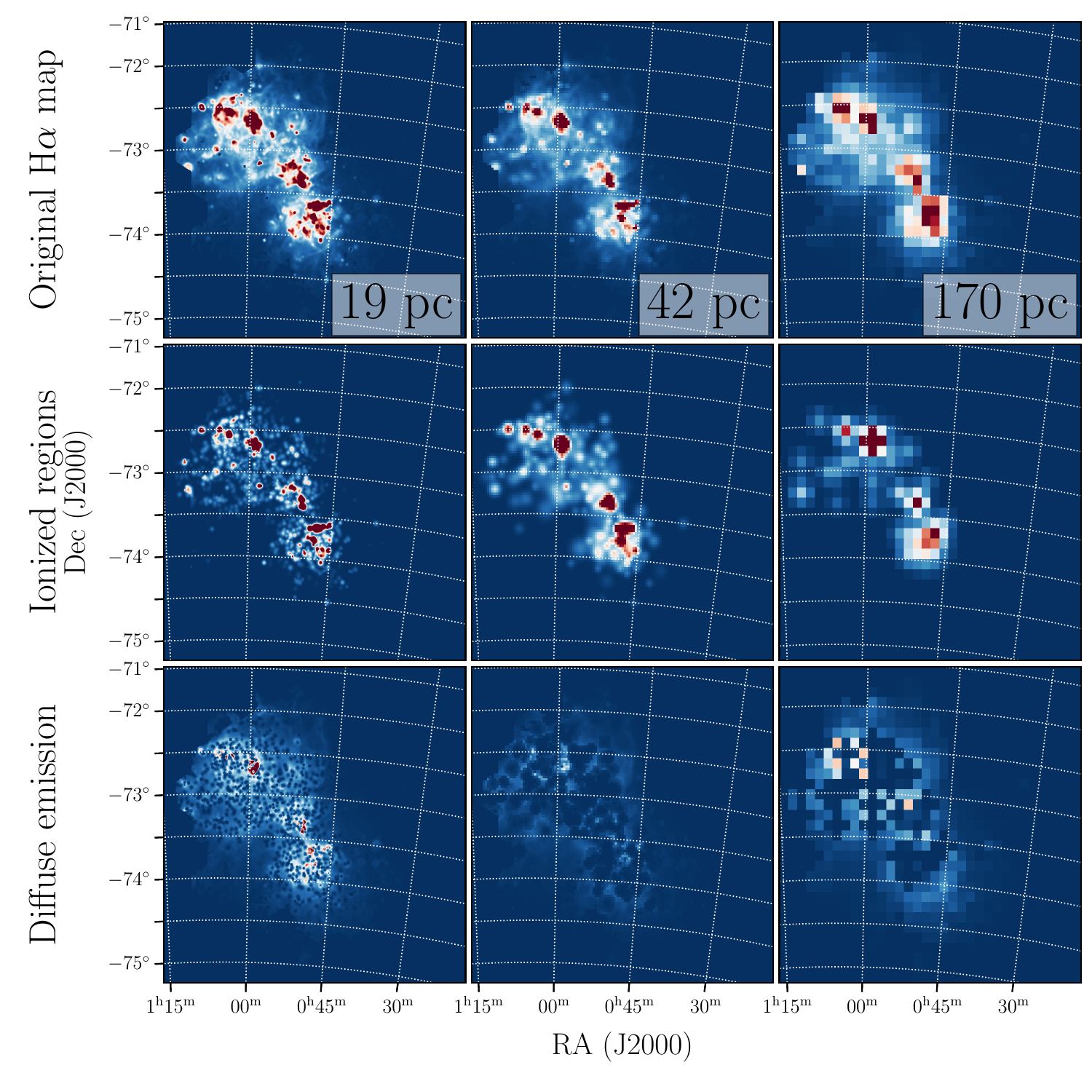}
    \caption{Spatial resolution analysis ($\sim$19 pc, $\sim$42 pc, and $\sim$170 pc, from left to right) of the SMC. Top row panels: Original H$\alpha$ emission intensity maps derived from the \texttt{pypipe3d} pipeline. Middle row panels: Ionized regions model generated with \texttt{pyHIIextractor}. Bottom row panels: Diffuse emission intensity maps building with the external pixels to high H$\alpha$ emission.}
    \label{fig:smc_pyhii}
\end{figure*}

\begin{figure*}
    \centering \includegraphics[width=0.99\textwidth]{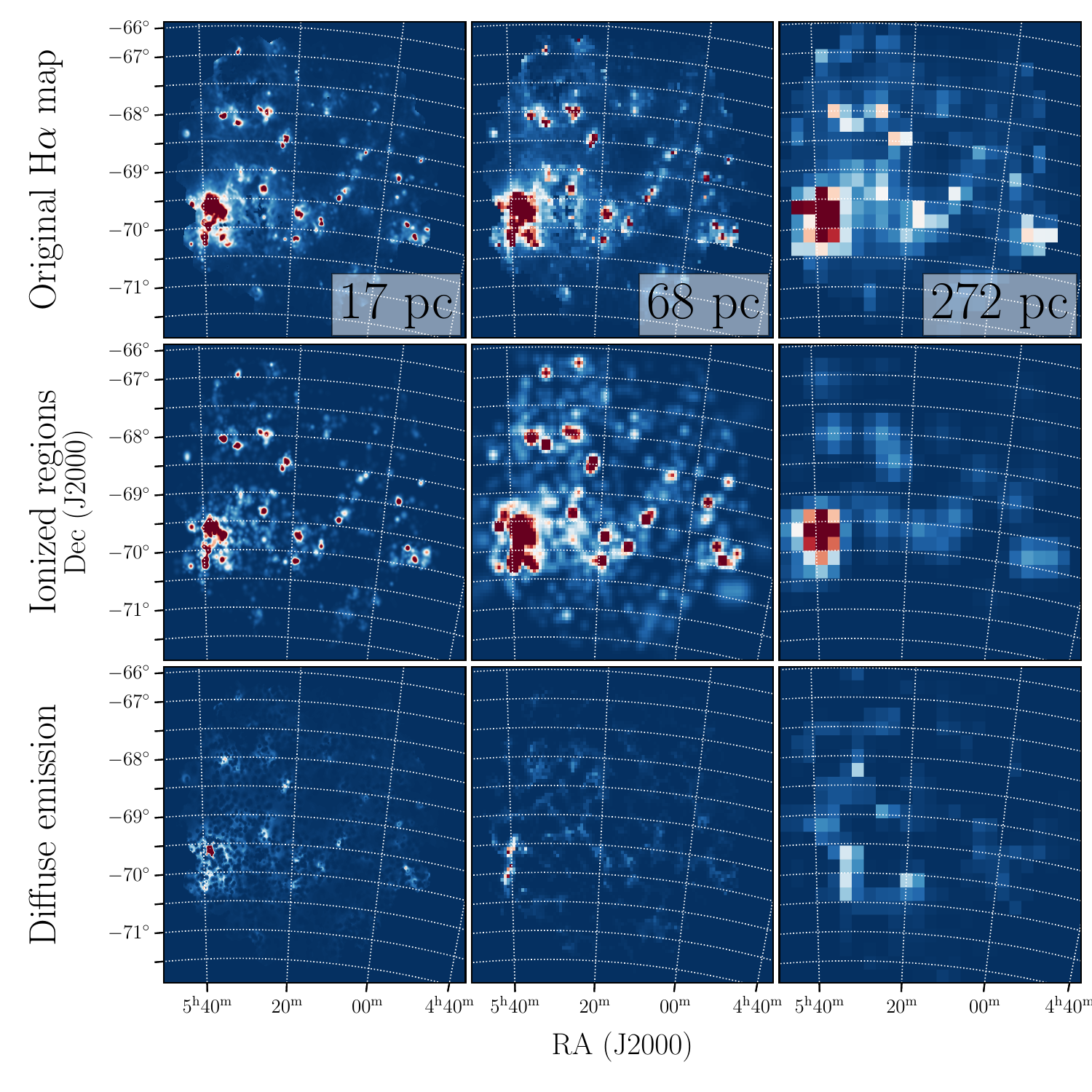}
    \caption{Spatial resolution analysis ($\sim$17 pc, $\sim$68 pc, and $\sim$272 pc, from left to right) of the LMC. The layout and description of the top, middle, and bottom panels are the same as in Figure \ref{fig:smc_pyhii}.}
    \label{fig:lmc_pyhii}
\end{figure*}

\begin{figure}
    \centering \includegraphics[width=0.4\textwidth]{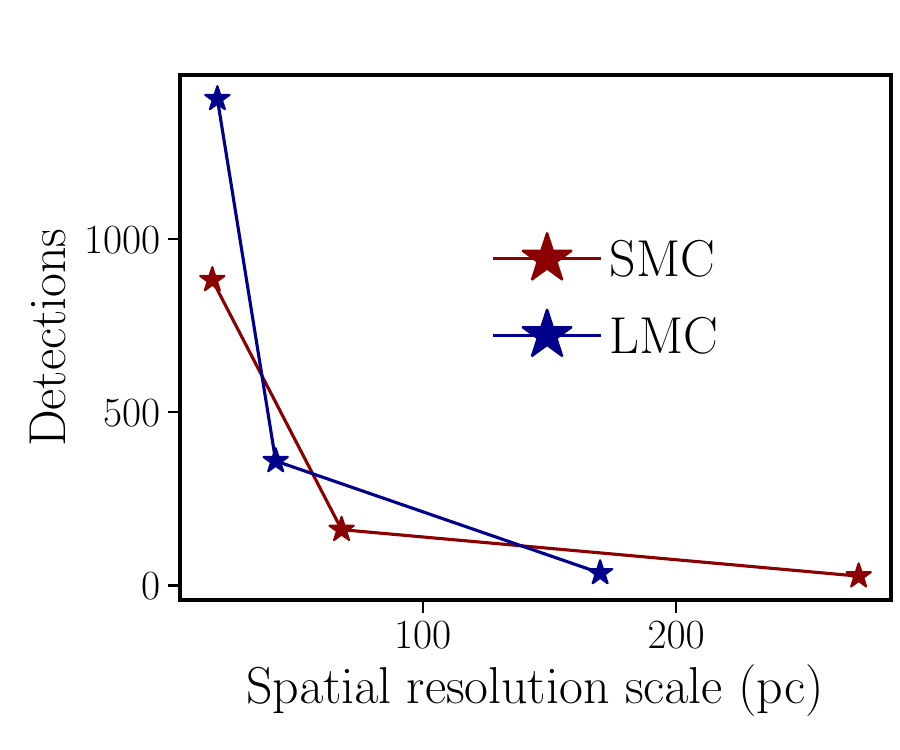}
    \caption{Clumpy region detections. Number of ionized regions identified by the \texttt{pyHIIextractor} code for the SMC and LMC (red and blue stars, respectively) as a function of spatial resolution (in parsecs).}
    \label{fig:detec}
\end{figure}

\begin{figure}
    \centering \includegraphics[width=0.4\textwidth]{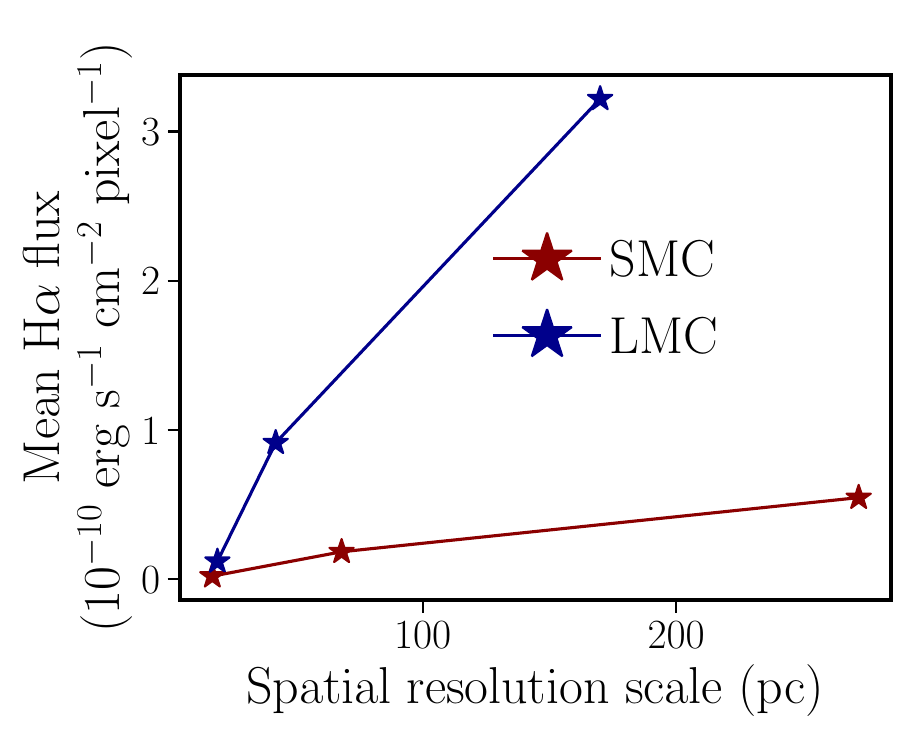}
    \caption{Mean integrated H$\alpha$ flux of all detected regions in each galaxy at each spatial resolution. Colors and labels are the same as in Figure~\ref{fig:detec}.}
    \label{fig:fluxes}
\end{figure}

\begin{figure}
    \centering \includegraphics[width=0.4\textwidth]{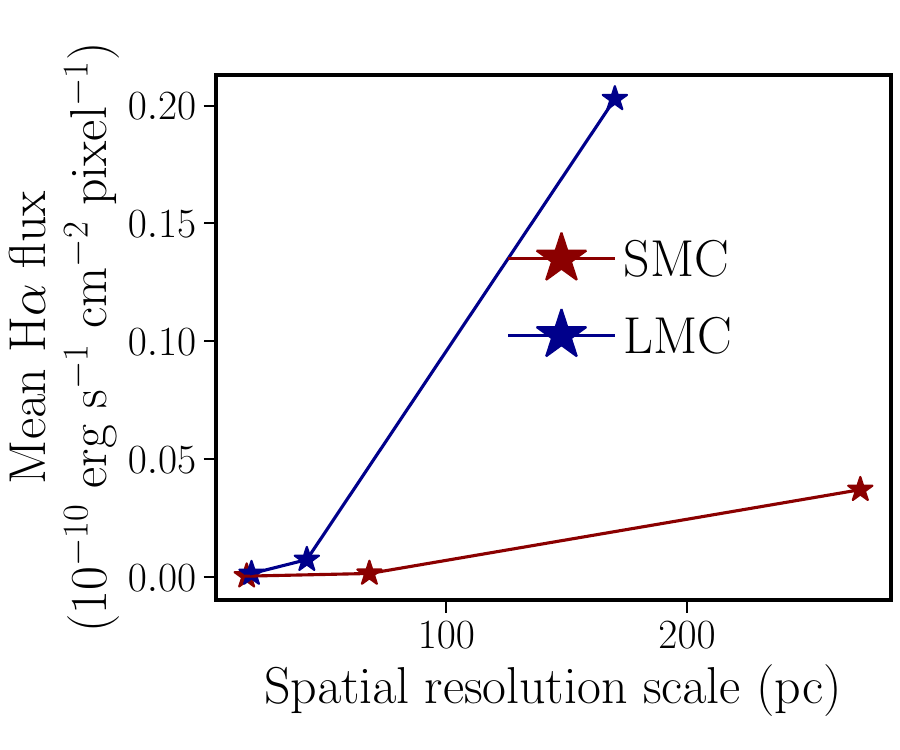}
    \caption{Mean integrated H$\alpha$ flux of diffuse emission at each spatial resolution. Colors and labels are the same as in Figure~\ref{fig:detec}.}
    \label{fig:fluxes_dig}
\end{figure}

Building the dataset for the Magellanic Clouds observed with the LVM at a 100\% fill factor requires nine dithered pointings per tile. The LMC observations were conducted in two periods: (i) from August 2023 to May 2024 (during the commissioning and early-science stage), and (ii) from July to November 2024 (during the early-science and survey-mode stages). Similarly, the SMC observations span two periods: (i) from October to November 2023 (during early science), and (ii) from July to November 2024 (during survey mode). Following the tile selection of Ibarra-Medel et al. (in prep.), a total of 1885 and 416 tiles are available for the LMC and SMC, respectively. After quality control, 1205 and 389 tiles are retained for the reconstruction of the data cubes used in this work.

Data reduction is performed using version 1.1.1 of the LVM Data Reduction Pipeline (LVM-DRP\footnote{\url{https://github.com/sdss/lvmdrp}}; Mejía-Narváez et al. (in prep.)), which is based on the Py3D package and adapted to the specific requirements of LVM data. The procedure includes homogenization of detector gains and correction for bias and cosmic rays; tracing and extraction of fiber spectra including stray-light correction; wavelength calibration and linear resampling; correction for fiber-to-fiber throughput variations; spectrophotometric calibration; combination of all channels into a single calibrated spectrum; sky modeling and subtraction; and implementation of the astrometric solution using guide-star images. The final product is a Row-Stacked Spectra (RSS) FITS file containing 1,944 spectra, accompanied by a table mapping the physical and astrometric positions of all fibers. 

\section{Methods}
\label{sec:methods}

The methodology followed for our analysis can be divided into three main steps: (i) the reconstruction of the datacubes; (ii) the measurement of emission-line properties; and (iii) the identification of fibers associated with ionized regions.

\subsection{Cube reconstruction}
\label{subsubsec:cube_rec}

For the reconstruction of a datacube for each galaxy, we use the RSS files produced by the DRP as input and follow the methodology described by Ibarra-Medel et al. (in prep.), which is initially based on the cube reconstruction presented by \cite{sanchezgalbanywalcher2023}. Briefly, the procedure is based on the Python package 3DCUBEGEN\footnote{\url{https://github.com/hjibarram/3DCubeGen}}, which has been extensively tested and applied in multiple projects, including the Calar Alto Legacy Integral Field Area Survey (CALIFA; \citealp{sanchezgalbanywalcher2023}), the Mapping Nearby Galaxies at Apache Point Observatory Survey (MaNGA; \citealp{ibarramedelavilasanchez2019}), and various studies based on MEGARA observations \citep{benitezibarramedelnegreete2023}.

The cube reconstruction process can be summarized in three main steps. (i) A 2D spatial interpolation of the flux contained in each fiber (previously selected for each spaxel and constrained by distance to improve computational efficiency) is performed for each dither and tile. This step includes an aperture correction and a kernel-based weighting scheme, yielding a flux estimate that optimally incorporates the average signal from adjacent fibers. (ii) A subsequent kernel selection is applied. (iii) Finally, a 2D spatial interpolation is carried out for each spectral pixel.

\subsubsection{Spatial-resolution degradation}
\label{subsubsubsec:spatial}

We therefore reconstruct the LVM observations at three spatial resolutions for each Magellanic Cloud, adopting distances of 49.97 kpc for the LMC and 62.44 kpc for the SMC, as described by Ibarra-Medel et al. (in prep.). The spatial resolutions are obtained by varying the kernel used during the two-dimensional interpolation of the original RSS data. Thus, each resolution is independently reconstructed from the original observations rather than obtained by smoothing an existing cube.

The spatial resolution of the reconstructed cube is controlled by the kernel scale $\sigma$, which determines the spatial extent over which the flux from neighbouring fibres contributes to a given output spaxel. Increasing $\sigma$ broadens the spatial kernel and consequently mixes the emission from structures over progressively larger physical scales. Thus, larger kernel sizes produce lower spatial resolution, allowing us to mimic the loss of spatial information that occurs when observations of nearby galaxies are analysed at increasingly coarser physical scales. The smallest adopted kernel corresponds to the highest spatial resolution that can be reliably recovered from the LVM data.

Although the original LVM point spread function (PSF) has a small full width at half maximum (FWHM) of $\sim$0.6'', the maximum achievable spatial resolution is limited by the fibre diameter of 35.3''. Following \citet{sanchezgalbanywalcher2023}, we adopt a minimum kernel size of 8.8'' to avoid losing spatial information, with a spaxel size of $0.75\sigma$. The smallest adopted kernel therefore corresponds to the highest spatial resolution that can be reliably recovered from the LVM data.

The different spatial resolutions ($\sim$19, $\sim$42, and $\sim$170 pc for the SMC; $\sim$17, $\sim$68, and $\sim$272 pc for the LMC) were obtained using kernel sizes ranging from 18.8' to 35.3'' and spaxel sizes spanning 14.1' to 26.4''. This configuration yields PSF FWHM values between 44.3' and 2.4' (see Table \ref{tab:lvm_kernels}). The adopted kernels follow the reconstruction configurations described by Ibarra-Medel et al. (in prep.). For each galaxy, we selected three configurations with successive kernel sizes differing by approximately a factor of four: corresponding to configurations f, d, and b for the SMC, and e, c, and a for the LMC, as identified by their labels (ID) in Table \ref{tab:lvm_kernels}. This selection provides a systematic sampling of spatial scales to investigate how spatial degradation affects the identification and characterization of ionized regions and their distribution in diagnostic diagrams.

\begin{table*}
\centering
\begin{tabular}{ccccc}
\hline
ID & Kernel size & Spaxel size & FWHM PSF & Spatial resolution (pc) \\
   &             &             &           & LMC \quad SMC \\
\hline
a & $18.8'$      & $14.1'$ & $44.3'$     & 272.88 \quad 340.98 \\
b & $9.4'$       & $7.1'$  & $22.1'$     & 136.44 \quad 170.49 \\
c & $4.7'$       & $3.5'$  & $11.0'$     & 68.22 \quad 85.25 \\
d & $2.3'$       & $1.7'$  & $5.4'$      & 34.11 \quad 42.62 \\
e & $1.2'$       & $54.0''$ & $2.8'$     & 17.05 \quad 21.31 \\
f & $35.3''$     & $26.4''$ & $2.4'$     & 15.22 \quad 19.08 \\
\hline
\end{tabular}
\caption{Kernel configurations adopted for the reconstruction of datacubes from the nine-dither LVM observations. The physical spatial resolution is derived from the PSF dispersion and computed using the adopted distances to the LMC and SMC. The physical spatial resolution values are taken from
Ibarra-Medel et al. (in prep.).}
\label{tab:lvm_kernels}
\end{table*}

\subsection{Emission-Line Property Measurements}
\label{subsubsec:dap}

Once the datacube is built, its spectral properties are analyzed in two complementary ways: (i) by fitting the nebular emission lines with Gaussian profiles as done by \citet{sattlermendezkreckel2025}, and (ii) by using the LVM-DAP\footnote{\url{https://github.com/sdss/lvmdap}} to measure the nebular emission-line fluxes in a parametric and non-parametric manner and to estimate the properties of the underlying stellar population. The rationale for using both types of measurements is to maintain strict control over the signal-to-noise ratio in the main CELs ([OIII]$\lambda$5007; [NII]$\lambda$6584; [SII]$\lambda\lambda$6717,31; and [OI]$\lambda$6300) and, in particular, to carefully assess potential sky contamination in [OI]$\lambda$6300, both in the line itself and in the continuum window from which the line noise is estimated. In principle, the velocity shift of the emission associated with the Magellanic Clouds allows a clear separation between nebular and sky emission; however, sky features may occasionally fall within the continuum windows where the DAP determines the line noise, leading to an overestimation.

The need to use the LVM-DAP is clear: it performs a more refined analysis to model the stellar continuum contribution, allowing us to estimate parameters such as the equivalent width, EW(H$\alpha$), the velocity dispersion, and the stellar absorption that may affect nebular lines such as those of HI, as described in detail in \citet{sanchezmejiaegorov2025}. The primary objective of the LVM-DAP is to decouple the stellar and ionized-gas components from the spectra observed by the LVM survey, and it operates following the philosophy of \texttt{pyFIT3D}, which in turn builds upon \texttt{pyPipe3D} \citep{lacerdasanchezmejia2022}. Both tools have been extensively applied to large-scale data analyses (e.g. CALIFA and MaNGA) and to simulations \citep{canosanchezzibetti2016, barreraheckmansanchez2018, ibarramedelavilasanchez2019}.

The DAP analysis begins with a nonlinear fit to determine the stellar systemic velocity (V${\star}$), velocity dispersion ($\sigma{\star}$), and dust attenuation (A$_{V\star}$). The stellar continuum is modeled using a linear combination of a simplified set of resolved stellar population synthesis (RSP) templates, adjusted for the instrumental resolution and line-spread function. A brute-force search identifies the best-fitting parameters while masking the bright emission lines. Stellar decomposition is performed only if the signal-to-noise ratio (S/N) at the normalization wavelength exceeds a user-defined threshold (default 20); for $1 < \mathrm{S/N} < 20$, a single best-matching template is adopted, and for $\mathrm{S/N} < 1$, no stellar model is created. Two modifications relative to \texttt{pyPipe3D} are implemented within the DAP analysis. First, the reconstructed datacube can be spatially binned to improve the signal-to-noise ratio for the stellar kinematic measurements and to match the spatial sampling adopted for the stellar population templates. Second, coarse spatial binning with smoothing is applied for the dust attenuation analysis to reduce the computational cost without affecting the results. These binning procedures are performed after the spatial degradation described above and do not define the spatial scales investigated in this work.

After determining the nonlinear parameters, a preliminary stellar model is subtracted, and strong emission lines are fit using Gaussian functions. Nearby lines may be grouped to tie their kinematic parameters and reduce the number of free parameters \citep{sanchezperezsanchez2016}. The fitting procedure is carried out in two steps: a pseudorandom search to identify the global minimum (the RND method), followed by a Levenberg–Marquardt minimization for precise parameter estimation. The parametric fit assumes a near-zero residual background after subtraction of the stellar continuum.

Stellar-component synthesis is performed on the residual continuum after emission-line subtraction, using the full RSP libraries constructed from the CoSHa-MaStar stellar spectra \citep{mejiabruzualsanchez2022}. These templates are shifted in velocity, broadened to match the stellar velocity dispersion, and attenuated for dust extinction using the previously derived nonlinear parameters. The synthesis yields (i) the coefficients of the linear combination of templates, and (ii) luminosity-weighted averages of the stellar physical parameters. The residuals are then smoothed to ensure an approximately zero-background spectrum, thereby facilitating an accurate emission-line fitting.

The stellar-subtracted (gas-only) spectrum is subsequently analyzed using both parametric and non-parametric methods. Parametric fitting provides measurements of the flux, velocity, velocity dispersion, and equivalent width of the strong emission lines, whereas non-parametric fitting uses weighted moments to analyze 192 predefined lines. Instrumental resolution corrections are not applied to the velocity dispersions, and spectral defects are masked using information from the DRP. Uncertainties are estimated through Monte Carlo iterations.

In this work, we reconstruct the datacube and subsequently analyze it; therefore, the inputs and outputs of the LVM-DAP were adapted so that the resulting products consist of two-dimensional maps of the properties of both, ionized gas and stellar component. Subsequently, a S/N $>$ 3 cut in the H$\alpha$ emission is applied to construct all RGB maps for both galaxies. Due to this criterion and the ability to obtain measured fluxes from different emission lines, each emission map is constructed using a different number of data points.

In Figure \ref{fig:MC_RGB} we present reconstructed RGB images of the Magellanic Clouds at different spatial resolutions, where the red, green, and blue channels correspond to the flux intensities of [NII]$\lambda6584$, H$\alpha$, and [OIII]$\lambda5007$, respectively. The upper panels display the SMC at physical scales of $\sim$19, $\sim$42, and $\sim$170 pc, while the lower panels show the LMC at $\sim$17, $\sim$68, and $\sim$272 pc.

At the finest scales (17–19 pc), both galaxies exhibit a clumpy morphology dominated by compact ionized regions surrounded by diffuse ionized gas (DIG) due to other physical process. The bright ionized regions are sharply resolved, particularly in the LMC, where several knots of intense emission are distinguishable. At intermediate scales (42–68 pc), individual ionized complexes start to blend into larger structures, while the diffuse emission becomes more prominent. Finally, at the coarsest resolutions (170–272 pc), the galaxies appear much smoother: compact regions are unresolved, and the emission is dominated by large-scale diffuse components, which retain the global distribution of ionized gas but wash out the small-scale substructure.

\subsection{Detection and Identification of Clumply Regions}
\label{subsec:pyHIIextractor}

After the H$\alpha$ emission-line maps are obtained from the DAP output at the various spatial resolutions considered in this work (SMC: $\sim$19, $\sim$42, and $\sim$170 pc; LMC: $\sim$17, $\sim$68, and $\sim$272 pc), they are analyzed using the \texttt{pyHIIextractor}\footnote{\url{https://github.com/sfsanchez72/pyHIIExplorer}} code to identify the ionized component. The following sections summarize the workflow of the code, including region detection and modelling.

\texttt{pyHIIextractor} is an open-access Python tool introduced by \citet{lugosanchezespinosa2022}, designed to identify ionized regions and extract their spectroscopic properties while correcting for contamination by the diffuse component, using high–spatial resolution IFS data \citep{lugosanchezbarrera2024}. It was initially conceived for datasets such as those from the All-weather MUse Supernova Integral-field Nearby Galaxies++ compilation (AMUSING++; \citealp{galbanyandersonrosales2016, lopezsanchezanderson2020}) obtained with the MUSE spectrograph, which delivers an FWHM–PSF of $\sim$0.8'' \citep{baconconseilmary2017}. The code extends and updates the algorithms implemented in \texttt{HIIExplorer} and \texttt{pyHIIexplorer} (\citealp{sanchezrosalesmarino2012}; \citealp{espinosasanchezmorisset2020}, respectively), which were previously applied to data from the PINGS survey \citep{rosaleskennicuttsanchez2010}, CALIFA \citep{sanchezperezrosales2015}, and MUSE \citep{sanchezsanchezperez2018}.

The program consists of two main stages: the identification of ionized regions and the extraction of emission-line and stellar population properties of each region. Ionized regions are defined as centrally concentrated, high-contrast structures relative to the diffuse ionized gas background. An important improvement over \texttt{HIIexplorer} is the capability to model both the ionized regions and the DIG, allowing for the decontamination of the extracted properties.

The first block of the code detects ionized regions in the H$\alpha$ map and generates separate models for these regions and the DIG. Detection is based on a modified \texttt{BLOB\_LOG} algorithm from \texttt{scikit-image} package \footnote{\url{https://scikit-image.org/}}, while the modelling assumes Gaussian profiles and applies a Laplacian of Gaussian (LoG) filter over a user defined range of dispersions ($\sigma$) linked to the physical size of the clumps.

The BLOB LoG algorithm stacks Laplacian-of-Gaussian filtered images derived over a range of $\sigma$ values into a three-dimensional space and identifies local maxima across spatial and scale dimensions. These maxima define both the location of intensity peaks and the characteristic size of the detected structures.

The DIG model is generated using a two-step approach. First, \texttt{pyHIIextractor} identifies the clumpy ionized regions using the BLOB/LoG algorithm on the H$\alpha$ emission-line map. A Delaunay triangulation is then constructed from the positions of these regions, and the points located farthest from the clumpy regions are selected as DIG tracers. Their fluxes are interpolated to construct an initial DIG map, which is subsequently subtracted from the emission-line image. As a second refinement of the initial DIG model, an optional but recommended step is applied to obtain a more consistent reconstruction of the diffuse emission. In this step, Gaussian models of the clumpy regions are subtracted from the original map to obtain a refined DIG image, at a higher computational cost. The diffuse flux at each pixel is then estimated by co-adding nearby tracer points within one FWHM, applying positive weights to DIG tracers and negative weights to clumpy regions, thereby producing an improved DIG reconstruction.

The second block of \texttt{pyHIIextractor} extracts spectroscopic properties of the detected clumps and is compatible with multiple data products and IFS analysis frameworks, including \texttt{pyPipe3D}, LVM-DAP, \texttt{pyCASSO} \citep{deamorimgarciacid2017}, and the MaNGA DAP \citep{westfallcappellariibershady2019}.

In summary, \texttt{pyHIIextractor} provides a practical way to identify clumpy ionized regions and separate them from the surrounding DIG, which is essential for reliable line-ratio analysis at different spatial resolutions. For each spatial resolution and for both Magellanic Clouds, the H$\alpha$ maps are analyzed to define the apertures of the ionized regions and extract their spectroscopic properties, including emission-line fluxes, EW(H$\alpha$), and H$\alpha$ velocity dispersion. In each case, the minimization of the $\chi^{2}$ parameter is used to determine the optimal flux thresholds for region detection. 

Figures \ref{fig:smc_pyhii} and \ref{fig:lmc_pyhii} present the maps of the Small and Large Magellanic Clouds, respectively, for: (i) the case of all pixels (upper panels), (ii) the ionized regions selected with \texttt{pyHIIextractor} or clumps in H$\alpha$ (middle panels), and (iii) the DIG emission outside the ionized regions (bottom panels). In general, the H$\alpha$ emission of both galaxies preserves the structure described in the RGB maps at all resolutions. 

\texttt{pyHIIextractor} detects clumpy regions in a manner comparable to how they would be identified at intermediate redshifts (0.05 < z < 0.5; \citealp{lugosanchezbarrera2024}). To first order, this segmentation provides a representative sampling of the ionized gas distribution in the present analysis. Detailed studies of the morphology of individual HII regions, including those ionized by single stars and stellar clusters, are the subject of ongoing work (Wofford et al. in preparation; Sánchez-Gallego et al. in preparation).

Figure \ref{fig:detec} shows the number of ionized regions identified by \texttt{pyHIIextractor} in the SMC (red stars) and LMC (blue stars). Figure \ref{fig:fluxes} presents the mean integrated H$\alpha$ flux of all detected regions for each galaxy, while Figure \ref{fig:fluxes_dig} shows the corresponding mean integrated H$\alpha$ flux of the external pixels and of the diffuse emission model. The DIG model produced by \texttt{pyHIIextractor} is shown here for descriptive purposes only and is not further explored or included in the analysis presented in this work. A separate study currently in preparation will investigate and compare different DIG-modelling techniques in the Magellanic Clouds using LVM data. This separation of the DIG analysis from the present work is consistent with our main goal, which is to characterize the diagnostic diagrams of the ionized regions identified by \texttt{pyHIIextractor}, without including the DIG component.

\section{Diagnostic diagrams}
\label{sec:diagnostic_diags}

In this work, we explore six different diagnostic diagrams, for which the following nomenclature will be adopted throughout the article:
O3 $\equiv$ log([OIII]$\lambda$5007/H$\beta$),
N2 $\equiv$ log([NII]$\lambda$6584/H$\alpha$),
S2 $\equiv$ log([SII]$\lambda$6716+30/H$\alpha$), 
O1 $\equiv$ log([OI]$\lambda$6300/H$\alpha$), 
n2 $\equiv$ [NII]$\lambda$6584/H$\alpha$, and
s2 $\equiv$ [SII]$\lambda$6716+30/H$\alpha$.
Using these definitions, the diagrams considered are:
\begin{enumerate} 
\item O3 versus N2 (classical BPT diagram)
\item O3 versus S2
\item O3 versus O1
\item s2 versus n2
\item log(EW(H$\alpha$)) versus N2, corresponding to the WHAN diagram
\item log(EW(H$\alpha$)) versus log($\sigma_{\mathrm{H}\alpha}$), corresponding to the WHaD diagram.
\end{enumerate}

These diagrams have been described by \citet{baldwinphillipsterlevich1981}, \citet{veilleuxosterbrock1987}, \citet{haffnerreynoldstufte1999}, \citet{cidfernandesstasinskaschlickmann2010}, and \citet{sanchezlugosanchezalmeida2024}, respectively.

\subsection{The O3–N2 Diagnostic Diagram}

\cite{baldwinphillipsterlevich1981} compiled a heterogeneous spectroscopic dataset encompassing HII regions, planetary nebulae, shock-ionized galaxies, and galaxies powered by hard radiation fields (including Seyferts and LINERs). 

The physical basis of the BPT classification lies in the combined use of recombination and collisionally excited emission lines, which respond differently to the properties of the ionizing radiation field and the thermodynamic state of the gas. Hydrogen recombination lines (e.g., H$\alpha$, H$\beta$) primarily trace the total flux of ionizing photons above 13.6 eV and are only weakly dependent on gas temperature and metallicity, making them robust normalizers of the ionized emission \citep{osterbrock89}. In contrast, collisionally excited forbidden lines probe the hardness of the ionizing spectrum and the physical conditions of the gas. In particular, [OIII]$\lambda5007$ traces highly ionized species (O$^{2+}$) and is sensitive to photons with energies above 35 eV, becoming enhanced in regions ionized by very young, massive stars, active galactic nuclei (AGN), or shocks. Conversely, [NII]$\lambda6584$ traces partially ionized gas and is strongly sensitive to metallicity and to the shape of the ionizing spectrum, with elevated emission commonly associated with AGN, DIG, and evolved stellar populations \citep{bonbonmarziani2025}. As a result, the vertical axis of the BPT diagram, [OIII]$\lambda5007$/H$\beta$, and the horizontal axis, [NII]$\lambda6584$/H$\alpha$, are both strongly sensitive to the electron temperature of the gas. Because each ratio involves emission lines that are close in wavelength, the impact of dust reddening on the measured ratios is minimized. This temperature depends on several physical factors, including the hardness of the ionizing radiation field, the ionization parameter, and the gas metallicity. 

Subsequent demarcation curves have formalized this physical separation, including the theoretical maximum starburst limit derived from photoionization models \citep{kewleydopitasutherland2001}, the empirical separation based on large observational samples \citep{kauffmanheckmantremonti2003}, and more recent corrections accounting for the contribution of DIG \citep{espinosasanchezmorisset2020}. In this work, we adopt these three demarcation lines to distinguish star-forming regions from harder ionization sources.

\subsection{O3 versus S2 Diagnostic Diagram}

\cite{veilleuxosterbrock1987} introduced an alternative formulation of the BPT diagnostic by replacing the [NII]$\lambda6584$ line with the [SII]$\lambda\lambda6717,6731$ doublet, based on a sample of 264 objects ranging from HII regions in external galaxies to narrow-line radio galaxies. While preserving the same physical foundation based on the combined use of hydrogen recombination lines and collisionally excited forbidden transitions, this diagram provides enhanced sensitivity to low-ionization and non-photoionization processes. In contrast to [NII]$\lambda6584$, whose parent ion (N$^{+}$) has an ionization potential of 14.5 eV and therefore preferentially traces the lower-ionization zones within the H$^{+}$ zone, [SII]$\lambda\lambda6717,6731$ arises from S$^{+}$, which has a lower ionization potential (10.4 eV) and is preferentially produced in partially ionized regions and extended transition zones. As a result, the [SII]$\lambda\lambda6717,6731$/H$\alpha$ ratio more directly traces the presence and extent of low-ionization gas, DIG, and shock excitation, although the [SII] lines still require sufficiently high electron temperatures to be collisionally excited and therefore weaken at low electron temperatures. In this sense, the ratio is generally less affected by metallicity than [NII]$\lambda6584$/H$\alpha$. This enhanced sensitivity allows the O3–S2 diagram to more effectively separate classical HII regions from shock-dominated regions, particularly in regimes where the classical BPT diagram exhibits significant overlap. Demarcation curves derived from large observational samples and photoionization models (\citealp{veilleuxosterbrock1987, kewleygroveskauffmann2006}), together with more recent corrections accounting for the contribution of DIG \citep{espinosasanchezmorisset2020}, formalize this separation. In this work, we adopt these demarcation lines to robustly distinguish star-forming regions from harder and low-ionization excitation sources.

\begin{figure*}
    \centering \includegraphics[width=0.6\textwidth]{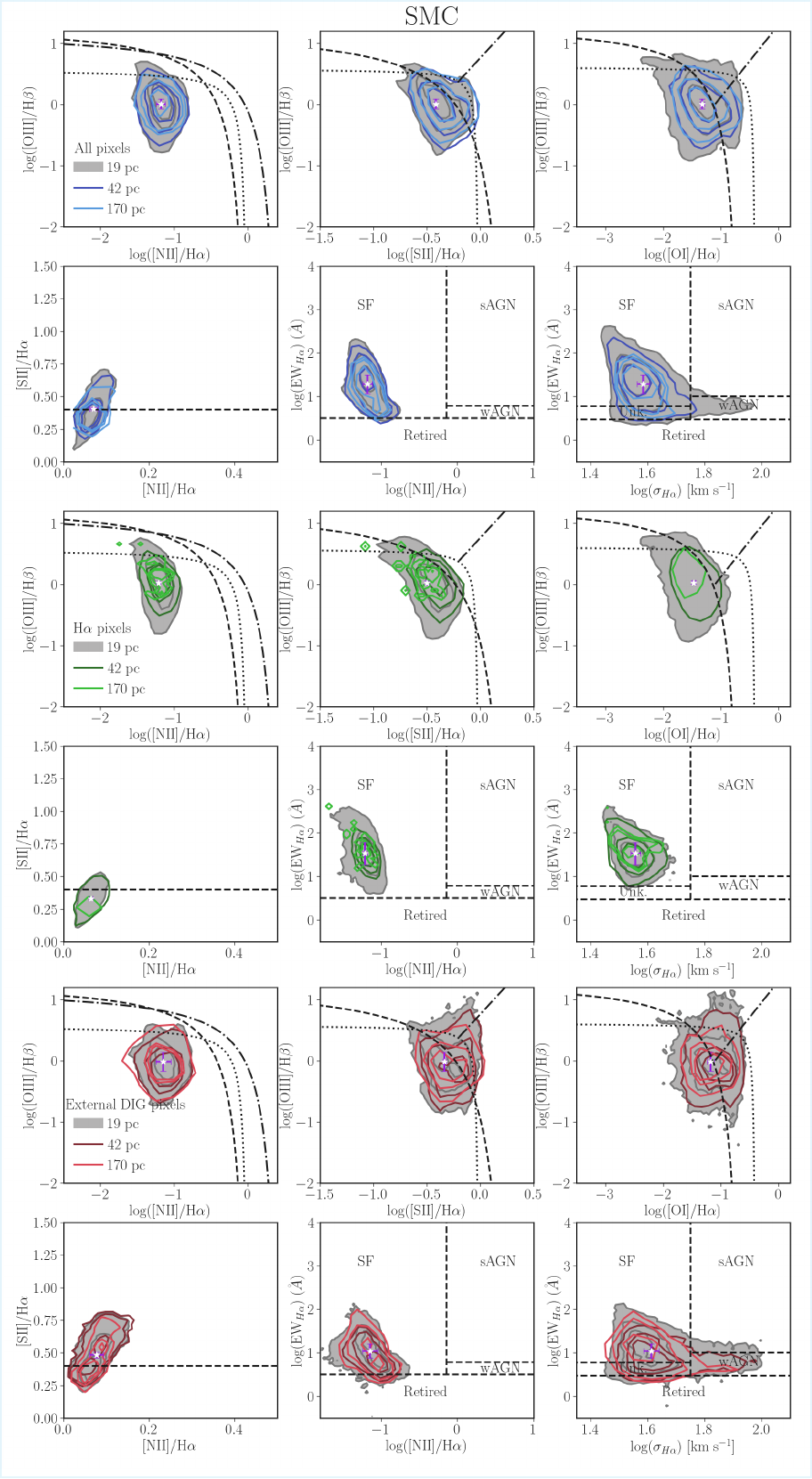}
    \caption{Diagnostic diagrams for the SMC. The three groups show distributions for all pixels in the map (top, blue contours), for those associated with high H$\alpha$ emission (middle, green contours), and those outside the high H$\alpha$ emission regions (bottom, red contours). Each panel displays the three different spatial resolutions $\sim$19 pc, $\sim$42 pc, and $\sim$170 pc as indicated in the upper left label. For each panel, in the upper row we show the O3 versus N2 (classic BPT diagram), O3 versus S2, and O3 versus O1 diagrams (from left to right); in the bottom row we show the [NII] versus [SII] diagram, the WHAN diagram, and the WHaD diagram (from left to right). The distributions in all panels are shown as density contours enclosing 99\%, 65\%, and 35\% of the regions, respectively. In cases where only a contour is shown, it corresponds to the 99\% level. In each: (i) classic BPT diagram, black dashed, dot-dashed and dotted lines represent \citet{kauffmanheckmantremonti2003}, \citet{kewleydopitasutherland2001} and \citet{espinosasanchezmorisset2020} demarcation lines, respectively; (ii) [OIII]$\lambda$5007/H$\beta$ versus [SII]$\lambda$6716 + 30/H$\alpha$ and [OIII]$\lambda$5007/H$\beta$ versus [OI]$\lambda$6300/H$\alpha$ diagnostic diagrams, black dashed (AGN region), dot-dashed (LINER/Seyfert type 2) and dotted lines (star formation) represent \citet{kewleygroveskauffmann2006} and \citet{espinosasanchezmorisset2020} demarcation lines, respectively; (iii) WHAN and WHaD diagrams show the regions associated with different ionization mechanisms, as defined by \citet{cidfernandesstasinskamateus2011} and \citet{sanchezlugosanchezalmeida2024}, respectively: star formation (SF), strong and weak AGN (sAGN, wAGN), and retired sources (Retired). Additionally, the WHaD diagram includes a region of unknown ionization (Unk.). In each panel, the white star represents the average value of each case, and the purple lines represent the propagated errors from the parameters used.}
    \label{fig:smc_diagrams}
\end{figure*}

\begin{figure*}
    \centering \includegraphics[width=0.6\textwidth]{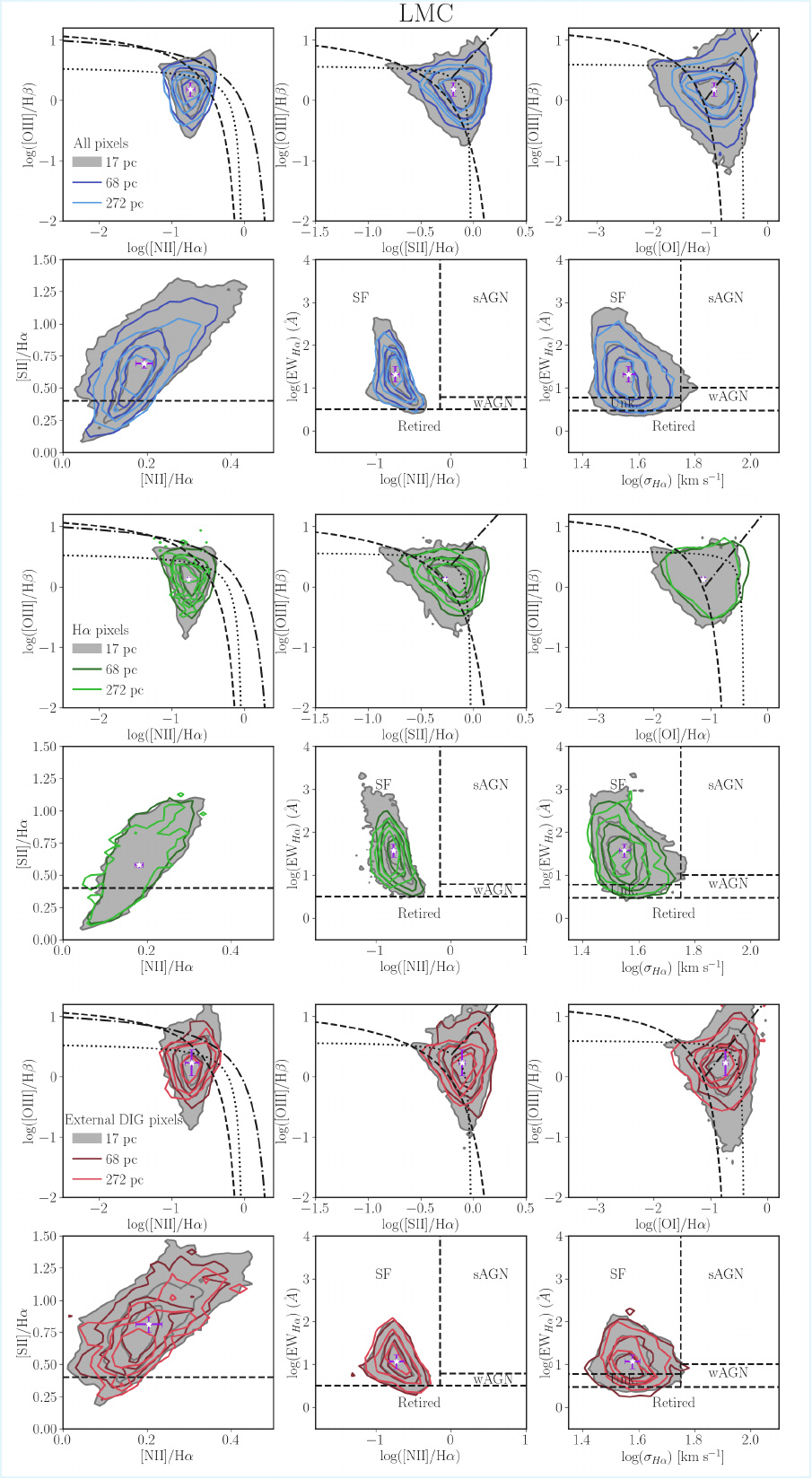}
    \caption{Diagnostic diagrams for the LMC. The description of the diagrams, the demarcation lines, colors, and contours is the same as in figure \ref{fig:smc_diagrams}. Each panel displays the three different spatial resolutions $\sim$17 pc, $\sim$68 pc, and $\sim$272 pc as indicated in the upper left label.}
    \label{fig:lmc_diagrams}
\end{figure*}

\begin{table*}
\centering
\begin{tabular}{lcccccc}
\hline
Case & N2 & O3 & S2 & O1 & log(EW(H$\alpha$)) & log($\sigma_{H\alpha}$) \\
\hline
SMC (i) & -1.1795 & 0.0086 & -0.4172 & -1.3242 & 1.2868 & 1.5842 \\
LMC (i) & -0.7418 & 0.1790 & -0.1908 & -0.9400 & 1.3190 & 1.5652 \\
SMC (ii) & -1.2117 & 0.0301 & -0.5005 & -1.4713 & 1.5290 & 1.5577 \\
LMC (ii) & -0.7642 & 0.1278 & -0.2718 & -1.1336 & 1.5681 & 1.5500 \\
SMC (iii) & -1.1467 & -0.0164 & -0.3343 & -1.1778 & 1.0444 & 1.6127 \\
LMC (iii) & -0.7234 & 0.2324 & -0.1097 & -0.7412 & 1.0681 & 1.5779 \\

\hline
\end{tabular}
\caption{Average values of the diagnostic parameters for the LMC and SMC considering all pixels, clumpy regions, and DIG regions.}
\label{tab:comparison_parameters}
\end{table*}

\subsection{O3 versus O1 Diagnostic Diagram}.

A more extreme formulation of the emission-line diagnostics replaces the low-ionization tracers [NII]$\lambda6584$ or [SII]$\lambda\lambda6717,6731$ with the [OI]$\lambda6300$ line, giving rise to the O3–O1 diagram. When a sufficiently high signal-to-noise ratio in [OI]$\lambda6300$ is available, this diagnostic is recommended as a powerful alternative \citet{veilleuxosterbrock1987}. The [OI]$\lambda6300$ forbidden line originates from neutral oxygen (O$^{0}$). Because the first ionization potential of oxygen (13.62 eV) is nearly identical to that of hydrogen (13.6 eV), neutral oxygen is confined to partially ionized or neutral regions where hydrogen is not fully ionized. In addition, charge–exchange reactions between hydrogen and oxygen, $\mathrm{H^{0} + O^{+} \leftrightarrow H^{+} + O^{0}}$, strongly couple their ionization states, further linking the presence of neutral oxygen to regions where hydrogen remains partially neutral. As a consequence, [OI]$\lambda6300$ emission arises predominantly in narrow transition layers between ionized and neutral gas, and shock-heated environments. The line is characterized by a relatively high critical density and a strong sensitivity to the electron temperature, and is primarily excited through collisional processes. In this framework, the vertical axis of the O3–O1 diagram, [OIII]$\lambda5007$, continues to trace the hardness of the ionizing radiation field, while the horizontal axis, [OI]$\lambda6300$/H$\alpha$, provides a stringent diagnostic of extended partially ionized zones and non-stellar excitation mechanisms, such as fast shocks and ionization like AGN-driven. This makes the O3–O1 diagram particularly effective at isolating LINER-like excitation and shock-dominated regions, with minimal contamination from classical HII regions.

\subsection{s2 versus n2 Diagnostic Diagram}

One of the earliest systematic uses of the [NII]$\lambda6584$/H$\alpha$ and [SII]$\lambda\lambda6717,6731$/H$\alpha$ line ratios to characterize the DIG was presented by \citet{haffnerreynoldstufte1999}, based on the WHAM H$\alpha$ Sky Survey of the Milky Way. By mapping ionized gas across different Galactic environments and heights above the mid-plane, this study demonstrated that low-ionization forbidden lines are systematically enhanced in the DIG relative to classical HII regions \citep{zhangyanbundy2017}. While sharing the same physical foundation as other emission-line diagnostics the [NII]$\lambda6584$/H$\alpha$ versus [SII]$\lambda\lambda6717,6731$/H$\alpha$ diagram is particularly sensitive to the structure, its density and extent of partially ionized gas. Both lines arise from collisionally excited low-ionization species, but probe different ionization regimes: N$^{+}$, with an ionization potential of 14.5 eV, closely follows the H$^{+}$ zone and makes [NII]$\lambda6584$/H$\alpha$ strongly sensitive to metallicity and moderately to the ionization parameter, whereas S$^{+}$, with a lower ionization potential of 10.4 eV \citep{wangheckmanamorin2021}, preferentially traces extended partially ionized regions and transition zones \citep{langergoldsmithpineda2016}. As a result, the [SII]$\lambda\lambda6717,6731$/H$\alpha$ ratio responds more directly to variations in electron density, ionization structure, and the presence of diffuse or shock-heated gas. Line emissivity models further show that variations in this diagram are primarily driven by changes in the ionized sulfur fraction (S$^{+}$/S), while the nitrogen ionization state remains approximately constant \citep{madsenreynoldshaffner2006}. Consequently, DIG-dominated regions systematically populate loci of elevated [SII]$\lambda\lambda6717,6731$/H$\alpha$ and [NII]$\lambda6584$/H$\alpha$, enabling a physically motivated separation between compact ionized regions and extended DIG. Observational and theoretical studies indicate that [SII]$\lambda\lambda6717,6731$/H$\alpha$ values above $\sim$0.4 provide a robust empirical threshold to identify DIG-dominated emission \citep[e.g.,][]{braccibelfioreginolfi2025}. However, this boundary depends on the gas metallicity and ionization conditions, and BPT-like diagnostics were originally calibrated for integrated regions. When applied to individual spaxels, this criterion should therefore be regarded as indicative rather than definitive. In this work, we adopt [SII]$\lambda\lambda6717,6731$/H$\alpha = 0.4$ as a practical threshold to distinguish between high- and low-ionization regimes, often associated with OB-star photoionized regions and DIG-dominated emission, respectively. This criterion is used purely as a line-ratio diagnostic and does not rely on the explicit identification of HII regions and DIG through different algorithms.


\subsection{WHAN Diagnostic Diagram}

Proposed by \citet{cidfernandesstasinskaschlickmann2010}, the WHAN diagnostic diagram is based on the combined use of the [NII]$\lambda6584$/H$\alpha$ line ratio and the EW(H$\alpha$), using a sample of $\sim$ 370,000 galaxies from the seventh data release of the Sloan Digital Sky Survey (SDSS). The EW(H$\alpha$) provides a direct measure of the ionizing photon production rate relative to the underlying stellar continuum. Because it normalizes the nebular emission by the local continuum, it is less sensitive to distance, aperture effects, and dust attenuation, making it a robust and observationally efficient diagnostic.

Following \citet{cidfernandesstasinskamateus2011}, a formal classification of five distinct ionization regimes is adopted: (i) pure star-forming galaxies (SF), characterized by $\log(\mathrm{[NII]}/\mathrm{H}\alpha)<-0.4$ and EW(H$\alpha$)$>3$ \AA; (ii) strong active galactic nuclei (sAGN), with $\log(\mathrm{[NII]}/\mathrm{H}\alpha)>-0.4$ and EW(H$\alpha$)$>6$ \AA; (iii) weak active galactic nuclei (wAGN), with $\log(\mathrm{[NII]}/\mathrm{H}\alpha)>-0.4$ and EW(H$\alpha$) between 3 and 6 \AA; (iv) and emission dominated by evolved stellar populations (retired systems), with EW(H$\alpha$)$<3$ \AA, whose ionization is dominated by old stellar populations; and (v) passive galaxies, characterized by very weak or absent line emission. For the purposes of this study, we adopt this classification within the WHAN diagram but do not explicitly consider the passive galaxy regime, focusing instead on the identification of retired systems.

\subsection{WHaD Diagnostic Diagram}

Recently introduced by \citet{sanchezlugosanchezalmeida2024}, the WHaD diagnostic diagram was developed using a large and diverse dataset comprising $\sim$ 9,000 galaxies from the MaNGA survey, $\sim$890 from the eCALIFA survey, and $\sim$40 AGNs from CALIFA. This extensive observational basis demonstrated that, regardless of the diagnostic employed, galaxies classified as star-forming are consistently identified by both classical BPT and WHaD schemes, while a significant fraction of X-ray–selected AGN are also recovered through optical diagnostics \citep{osoriogonzalezsanchez2023}. Moreover, the spatially resolved classification enabled by WHaD increases the sensitivity to harder, AGN-like ionization processes. In the WHaD diagram, star-forming sources occupy regions with high EW(H$\alpha$) values ($>6$ \AA) and low H$\alpha$ velocity dispersions ($\sigma_{\mathrm{H}\alpha}<57$ km s$^{-1}$). Retired sources are defined by low EW(H$\alpha$) values ($<3$ \AA), while weak and strong AGN occupy the higher-dispersion region ($\sigma_{\mathrm{H}\alpha}>57$ km s$^{-1}$) among non-retired regions and are separated according to their EW(H$\alpha$), with a boundary at EW(H$\alpha$) = 10 \AA. ``Unknown'' sources correspond to systems with intermediate EW(H$\alpha$) values and low velocity dispersions that cannot be securely classified as star-forming or AGN.

The WHaD diagram combines the EW(H$\alpha$), with the velocity dispersion of the H$\alpha$ line, thereby incorporating kinematic information into the classification of ionization sources. By relying exclusively on parameters derived from a single recombination line, this diagram minimizes uncertainties associated with multi-line measurements and signal-to-noise limitations, while retaining sensitivity to the dominant physical mechanisms governing the ionized gas. As in the WHAN formalism, EW(H$\alpha$) traces the ionizing photon production rate relative to the stellar continuum, providing a robust indicator of the strength and origin of the ionizing sources but with the inclusion of the H$\alpha$ velocity dispersion introduces a direct probe of the dynamical state of the gas.

In star-forming HII regions, line broadening is primarily governed by thermal motions, internal turbulence, and gravitational motions within the host galaxy, typically resulting in low velocity dispersions of $\sigma_{\mathrm{H}\alpha} \sim 15$–30 km s$^{-1}$ after correction for instrumental broadening. In contrast, shock-dominated regions, AGN-driven outflows, and mechanically heated gas often exhibit significantly enhanced dispersions, with $\sigma_{\mathrm{H}\alpha} \gtrsim 50$–100 km s$^{-1}$, reflecting the injection of kinetic energy by winds, jets, and large-scale feedback processes. Because shock heating and turbulent dissipation can also contribute to gas ionization, elevated velocity dispersions are naturally associated with harder and non-stellar excitation mechanisms.

The WHaD diagram therefore provides a physically motivated separation between photoionization by young massive stars and ionization driven by energetic processes, such as shocks, using only minimal observational inputs. Its ability to combine radiative and kinematic diagnostics enables an efficient identification of composite and AGN-like regions in spatially resolved data, particularly in regimes where classical line-ratio diagnostics become uncertain. In this work, we adopt the WHaD classification scheme of \citet{sanchezlugosanchezalmeida2024} to characterize the interplay between ionization and gas dynamics.

The six diagnostic diagrams described above probe different, yet complementary, aspects of the ionized gas, including ionization structure, excitation mechanisms, chemical enrichment, and kinematics. Together, they provide a general but no less important view of the physical conditions within the interstellar medium.

\begin{figure*}
    \includegraphics[width=0.45\textwidth]{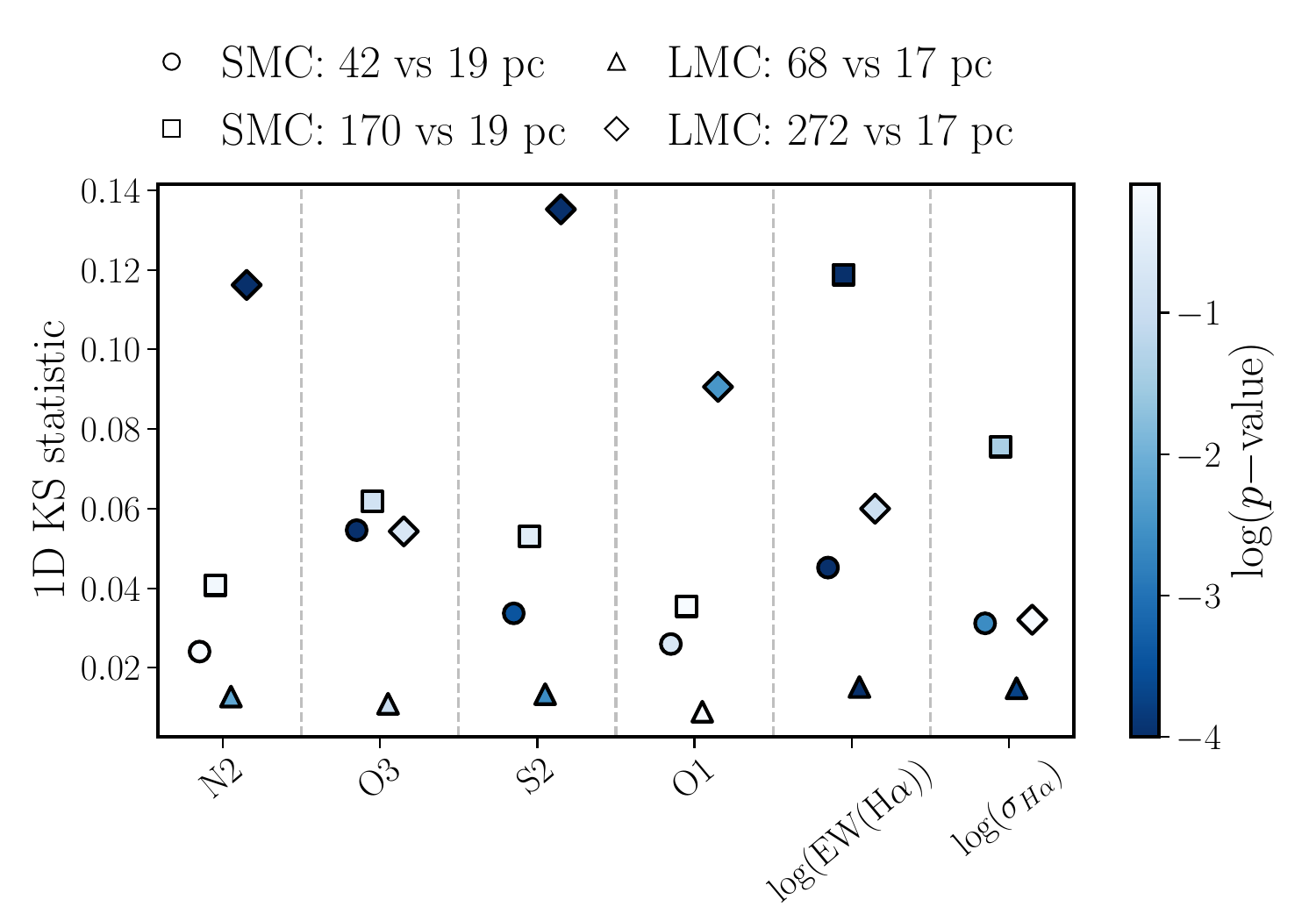}
    \includegraphics[width=0.45\textwidth]{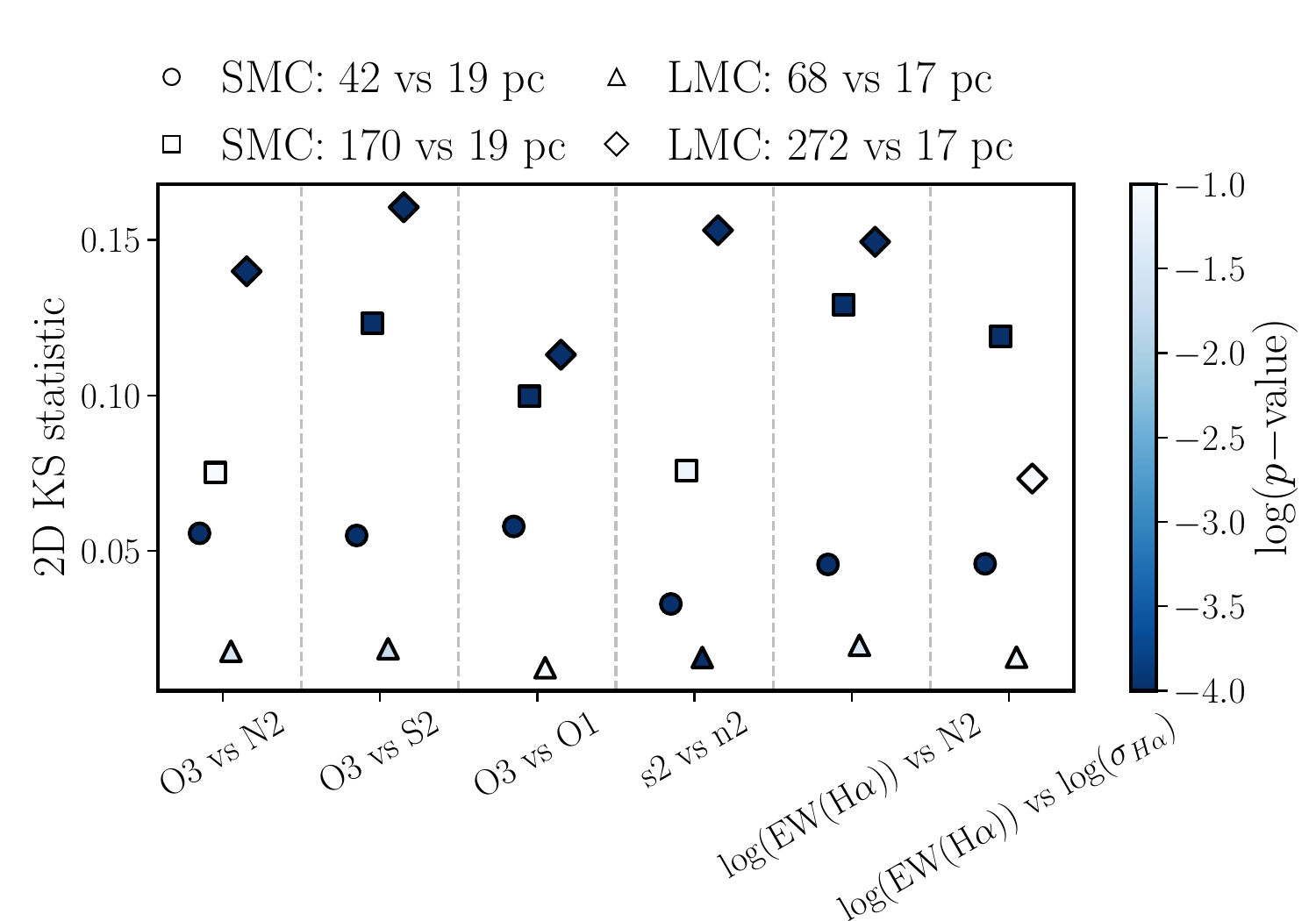}
    \caption{KS test results for the SMC and LMC for each parameter used in the diagnostic diagrams and for each diagnostic diagram. In both panels for the SMC, we compare $\sim$42 pc and $\sim$170 pc to the $\sim$19 pc baseline, while for the LMC we compare $\sim$68 pc and $\sim$272 pc to the $\sim$17 pc baseline. Left panel: 1D KS test where the columns correspond to N2, O3, S2, O1, EW(H$\alpha$) and $\sigma_{H\alpha}$. Right panel: 2D KS test where the columns correspond to O3 vs. N2, O3 vs. S2, \textbf{O3 vs. O1}, n2 vs. s2, WHAN, and WHaD diagrams. In both panels each symbol represents a different case, the color bar indicates the log(p-value) obtained and the dotted lines serve only to separate the different diagram cases.}
    \label{fig:KS}
\end{figure*}

\section{Results}

We adopt the same initial selection of pixels with a signal-to-noise ratio greater than 3 in the H$\alpha$ line for both galaxies. To remain consistent with the S/N cut applied in the construction of the RGB maps, this same threshold is used for pixel selection throughout the analysis. We prioritized a high signal-to-noise ratio, although we do not rule out the potential loss of information, particularly for regions with lower EW(H$\alpha$). The impact of this, along with the application of alternative methodologies for identifying low-ionization regions, will be explored in the future. In this work the effect of spatial resolution is explored in three different cases: (i) all pixels; (ii) ionized regions identified with the \texttt{pyHIIextractor} code; and (iii) regions dominated by DIG outside the ionized regions. The construction of each case is based on the H$\alpha$ map, as it represents the most intense emission line observed. 

\subsection{Detection of different phases of ISM}

Figures \ref{fig:smc_pyhii} and \ref{fig:lmc_pyhii} show that at the highest spatial resolution available for each galaxy ($\sim$19 pc and $\sim$17 pc), there are a larger number of detections with high H$\alpha$ emission (see Figure \ref{fig:detec}, SMC at $\sim$19 pc: 881, LMC at $\sim$17 pc: 1403), compared to those obtained at intermediate spatial resolution (SMC at $\sim$42 pc: 161, LMC at $\sim$68 pc: 359). At coarser resolutions, the number of detections decreases (SMC at $\sim$170 pc: 27, LMC at $\sim$272 pc: 36), leading to a loss of structural detail in each case. This implies that, at each spatial degradation scale considered (independently of the galaxy), we lose on average 82\% of the detections. On the other hand, the mean integrated H$\alpha$ flux of all detected regions increases by a factor of $\sim$7–11 during the first degradation in spatial resolution (see Figure \ref{fig:fluxes}, SMC: 19–42 pc; LMC: 17–68 pc) and by a factor of $\sim$3 during the second step (SMC: 42–170 pc; LMC: 68–272 pc), reflecting the progressive blending of individual regions and the DIG.

By comparing detections across consecutive spatial resolutions, we quantify the degree of blending. In the SMC, each low-resolution detection contains on average $\sim$3–4 and $\sim$12–13 high-resolution regions when degrading from $\sim$19 pc to $\sim$42 pc and $\sim$19 pc to $\sim$170 pc, respectively. Correspondingly, $\sim$65\% and $\sim$37\% of the $\sim$19 pc regions are blended into larger structures at $\sim$42 pc and $\sim$170 pc. In the LMC, each low-resolution detection contains on average $\sim$2–3 and $\sim$17–18 high-resolution regions when degrading from $\sim$17 pc to $\sim$68 pc and from $\sim$17 pc to $\sim$272 pc, respectively, with $\sim$60\% and $\sim$45\% of the $\sim$17 pc regions merged into larger structures. For both galaxies, while the average number of merged regions increases at coarser resolutions, the fraction of high-resolution regions associated with low-resolution detections decreases, indicating that a significant fraction of compact regions is lost into diffuse emission.

In the case of DIG maps (SMC and LMC) constructed from the pixels outside the clumpy regions (see bottom panels in Figures \ref{fig:smc_pyhii} and \ref{fig:lmc_pyhii}), it is visually apparent that at the highest spatial resolutions, enhanced inter-region or filamentary emission can be observed. When the resolutions are degraded and the sizes of the detections increase, these filamentary and inter-region structures are merged into the ionized regions, resulting in a greater loss of structural detail. Assuming that the diffuse emission at the highest resolution ($\sim$19 pc for the SMC and $\sim$17 pc for the LMC) represents the most realistic model of the DIG, the integrated H$\alpha$ flux of the DIG increases substantially with spatial degradation: by 5 times when degrading to $\sim$42 pc and $\sim$68 pc, respectively, and by 27 times at the coarsest resolutions of $\sim$170 pc and $\sim$272 pc for the SMC and LMC, respectively. These results highlight the strong impact of spatial resolution on both the morphology and the measured flux of the DIG.


\subsection{Diagnostic diagrams}

We explore six diagnostic diagrams for three pixel selections: i: all pixels, ii: H$\alpha$ pixels associated with ionized regions detected with \texttt{pyHIIextractor}, and iii: external DIG pixels. For each galaxy, the highest spatial resolution is compared with two lower resolutions. A summary of the results is shown in Figures \ref{fig:smc_diagrams} (SMC) and \ref{fig:lmc_diagrams} (LMC).

\subsubsection{SMC}

For case (i), considering all pixels, the O3–N2, O3–S2, and WHAN diagrams in Figure \ref{fig:smc_diagrams} show that the three density contours (99\%, 65\%, and 35\%) correspond to regions typically associated with star formation. In contrast, the O3-O1 diagram shows a slight shift of the contours towards higher O3 and O1 values, with part of the distribution overlapping the region above the standard demarcation lines. This may indicate a contribution from harder ionizing sources.

In the s2–n2 diagram, all contours are located within the ranges s2 $<$ 0.75 and n2 $<$ 0.1, covering from low to hard ionization. In the WHaD diagram, the contours for the three spatial resolutions cover different ionization zones, but in particular, the best resolution ($\sim$19 pc) reaches higher velocity dispersion values. 

In the case (ii), H$\alpha$ pixels, described by middle panels of Figure \ref{fig:smc_diagrams}, the contours corresponding to all spatial resolutions are similar to ionization star forming type, characterised by values of log(EW(H$\alpha$)) $>$ 0.7, where EW(H$\alpha$) is expressed in \AA, log($\sigma_{\mathrm{H}\alpha}$) $<$ 1.7, where $\sigma_{\mathrm{H}\alpha}$ is expressed in km s$^{-1}$, N2 $<$ –0.7, O3 $<$ 0.7, S2 $<$ –0.2, n2 $<$ –0.1, and s2 $<$ 0.5. Overall, the contours from the three spatial resolutions exhibit an even higher degree of overlap than in case (i), spanning very similar parameter ranges in all diagnostic diagrams. Nevertheless, in the outermost contours (99\%), pixels associated with strong H$\alpha$ emission at the highest resolution ($\sim$19 pc) remain dominant and provide more information compared to the other two spatial resolutions.

Finally, in case (iii), External DIG pixels, shown in the bottom panels of the Figure \ref{fig:smc_diagrams}, the contours corresponding to the highest spatial resolution no longer extend as broadly across the diagnostic diagrams as those in the H$\alpha$ pixel panels. In all diagrams, the contours encompass regions indicative of harder ionization than that expected from star forming regions, with this effect being particularly evident in the O3–S2, O3-O1, s2–n2, and WHaD diagrams. The density contours in case (iii) are characterized by low EW(H$\alpha$) values and, in the WHaD diagram, by a wide range of velocity dispersions, spanning all intervals associated with different ionization sources.

Overall, the highest spatial resolution achieved by LVM covers larger areas in all diagnostic diagrams for SMC, and as the spatial resolution decreases, only the central contours remain consistent. When comparing the three cases, the highest resolution also encompasses broader regions than the other two, revealing a certain homogenization of the data as the resolution is degraded. The most prominent differences between spatial resolutions arise mainly in the extreme values.

\subsubsection{LMC}

For case (i), all pixels shown in the upper panels of Figure \ref{fig:lmc_diagrams}, the density contours in the O3–N2 diagram exhibit high N2 values (N2 $>$ –1) compared to those observed in the SMC (N2 $<$ –1). The contours in the O3–S2 and O3–O1 diagrams indicate harder ionization, located near the intersection of the demarcation lines. In the s2–n2 diagram, the values span a notably wide range (0.01 $<$ s2 $<$ 1.4), extending beyond those typically associated with star-forming regions (s2 $\sim$ 0.4). Similarly, in the WHaD diagram, the contours show a broader dispersion and coverage; however, they remain predominantly within the regions associated with star formation and the “Unknown” classification. The contours show a clear variation in EW(H$\alpha$), while the velocity dispersion extends towards the AGN region, although over a somewhat narrower range than in the SMC. In contrast, in the WHAN diagram, all contours fall entirely within the star-forming zone.

The middle panels of Figure \ref{fig:lmc_diagrams} present case (ii), H$\alpha$ pixels, corresponding to pixels associated with strong H$\alpha$ emission. In the O3–N2, O3-S2 and O3-O1 diagram, improving the spatial resolution results in broader contours; however, only those at the highest resolution ($\sim$17 pc) reach the lowest O3 values. In both the O3–S2 and O3–O1 diagrams, the contours encompass regions spanning a wide range of ionization, from low values typical of star-forming regions to those characteristic of harder ionization, with the highest resolution providing the broadest coverage. In the s2–n2 diagram, the maximum contour values show a clear gradient, with higher s2 values attained at higher spatial resolution. In the WHAN diagram, the contours at all three spatial resolutions remain confined to ionization levels consistent with star-forming regions, with similar N2 values. In contrast, in the WHaD diagram, the contours extend across regions usually associated with star formation and the “Unknown” ionization classification.

The bottom panels of Figure \ref{fig:lmc_diagrams} present case (iii), External DIG pixels. In the O3–N2 diagram, the contours at different spatial resolutions cover the similar extent within the region corresponding to ionization like star formation and a harder ionization. By contrast, in the O3–S2 and O3–O1 diagrams, the contours at all resolutions show good agreement and extend into regions of harder ionization above the demarcation lines. In the s2–n2 diagram, higher values are reached at higher spatial resolution compared to case (ii) (s2 $>$ 1.25). In the WHAN and WHaD diagrams, the contours primarily cover low N2 values and low velocity dispersion, but they are mainly concentrated at low EW(H$\alpha$) values (log(EW(H$\alpha$)) $<$ 2).

\subsubsection{Comparison of Results from Diagnostic Diagrams}

Overall, the LMC spans a wider range of values in all diagnostic diagrams compared to the SMC. In particular, higher values of N2, S2, and O1 are reached, which may indicate the presence of regions with harder ionizing radiation fields or additional excitation mechanisms. However, the EW and velocity dispersion parameters remain predominantly within regions associated with star formation, in contrast to the case of the SMC. With the exception of the DIG case in the LMC, the contours at the three different spatial scales overlap in a more uniform manner.

The average values show systematic differences between the LMC and the SMC for all cases considered (all pixels, clumpy H$\alpha$ regions, and External DIG). In general, the SMC presents lower values of N2, S2, and O1, than the LMC, with mean N2 ranging from $\sim$-1.21 to -1.15 in the SMC compared to $\sim$-0.76 to -0.72 in the LMC. A similar offset is observed for S2 and O1, indicating systematically weaker low-ionization forbidden lines relative to H$\alpha$ in the SMC. These differences are consistent with the lower metallicity of the SMC. In contrast, the mean O3 values are comparable between the two galaxies, although the LMC tends to show slightly higher values in the DIG. When compared with the standard demarcation curves in BPT diagrams, the average line ratios for both galaxies lie well within the star-forming region regime and remain far from the "AGN'' or shock-ionized regions, particularly in the N2 and S2 diagrams. Additionally, the mean EW(H$\alpha$) values are above the typical DIG/retired boundary (log(EW(H$\alpha$))$\sim$3), especially in the clumpy regions, consistent with ionization dominated by massive stars. While the relative positions in the diagnostic diagrams remain similar, the systematic shift toward lower N2, S2, and O1 ratios in the SMC reflects its lower chemical enrichment compared to the LMC.

\subsection{Kolmogorov–Smirnov tests}

To assess the statistical consistency of the parameters and line ratios across different spatial resolutions, both one–dimensional (1D) and two–dimensional (2D) Kolmogorov–Smirnov (KS) tests are employed. The results are summarized in Figure \ref{fig:KS}, showing the 1D and 2D KS tests in the left and right panels, respectively. The 1D KS test is applied independently to each emission–line ratio used in the diagnostic diagrams, allowing the identification of systematic shifts or changes in their individual distributions between different spatial resolutions. In contrast, the 2D KS test, following the method of \cite{fasanofranceschini1987}, is used to evaluate the joint distributions of line ratios, thus, preserving the intrinsic correlations that define the structure of the diagnostic planes. The combined use of these approaches provides complementary insights: the 1D KS test offers robust and easily interpretable results for individual parameters, while the 2D KS test captures deviations in the global shape of the distributions, which are critical in the context of diagnostic diagrams. All KS tests, both 1D and 2D, are applied to all pixels (case i), allowing us to work with the statistically largest sample.

\subsubsection{SMC}

The left panel in Figure \ref{fig:KS} shows the 1D KS test results for the SMC, comparing multiple spatial resolutions ($\sim$42 pc and $\sim$170 pc) with the baseline value of $\sim$19 pc. At $\sim$42 pc, several parameters show statistically significant differences from the baseline: O3 and EW(H$\alpha$) exhibit the smallest p-values (p $<$ 10$^{-4}$), followed by S2 (p = 0.0002) and $\sigma_{H\alpha}$ (p = 0.0006). N2 and O1 show smaller but still significant differences (p-values between 0.0071 and 0.0157). At $\sim$170 pc, most parameters are statistically consistent with the baseline, with the highest p-values found for O1 (p = 0.8013), N2 (p = 0.6455), and S2 (p = 0.3191). The parameters that remain statistically different at this coarser resolution are $\sigma_{H\alpha}$ (p = 0.0489) and EW(H$\alpha$) (p = 0.0002), with the latter showing the strongest deviation.

The right panel in Figure \ref{fig:KS} shows the results of the 2D KS test. At $\sim$42 pc all diagnostic diagrams (O3 vs N2, O3 vs S2, O3 vs O1, s2 vs n2, WHAN, and WHaD) exhibit statistically significant differences (p $<$ 10$^{-4}$), indicating that even moderate degradation alters the joint distributions of diagnostic parameters. At $\sim$170 pc, strong deviations persist for several diagrams, particularly O3 vs S2, O3 vs O1, WHAN, and WHaD (all p $<$ 10$^{-4}$), while O3 vs N2 (p = 0.3000) and s2 vs n2 (p = 0.2000) show weaker or non-significant differences, suggesting some resilience to resolution changes. The largest KS values at $\sim$170 pc are observed in O3 vs S2 (0.1232) and WHAN (0.1292) diagrams.

\subsubsection{LMC}

For the LMC, in the 1D KS test (left panel of Figure \ref{fig:KS}) at $\sim$68 pc all parameters are largely consistent with the baseline ($\sim$17 pc), with the highest p-values for O1 (p = 0.8282), O3 (p = 0.5957), and N2 (p = 0.4041). No parameter shows a strongly significant difference at this resolution, although EW(H$\alpha$) and $\sigma_{H\alpha}$ have somewhat lower p-values (0.2148 and 0.2365, respectively). At $\sim$272 pc, several parameters show notable deviations: S2 exhibits the smallest p-values (p $<$ 10$^{-4}$; KS = 0.1352), followed by N2 (p = 0.0001; KS = 0.1162) and O1 (p = 0.0034; KS = 0.0906). The most stable parameters at this coarser resolution are O3 (p = 0.1992; KS = 0.0543) and $\sigma_{H\alpha}$ (p = 0.8109; KS = 0.0321), while EW(H$\alpha$) shows a moderate difference (p = 0.1197; KS = 0.0600).

For the 2D KS test (right panel of Figure \ref{fig:KS}), at $\sim$68 pc, most diagnostic diagrams exhibit no significant differences (p-values $>$ 0.1000), indicating that mild degradation has a limited effect on the joint distributions. The exception is s2 vs n2, which exhibits a significant deviation even at this resolution (p $<$ 10$^{-4}$). At $\sim$272 pc, all diagrams except WHaD show strong and systematic deviations (p $<$ 10$^{-4}$), with the largest KS values found for O3 vs S2 (KS = 0.1605), s2 vs n2 (KS = 0.1531), and WHAN (KS = 0.1494). By contrast, WHaD remains relatively stable (p = 0.1000; KS = 0.0733), confirming its lower sensitivity to spatial resolution.

\subsubsection{Comparison of KS Test Results}

Thus, for 1D KS tests, in the SMC, the parameters most sensitive to spatial resolution are O3, EW(H$\alpha$), and S2, which show statistically significant differences even at intermediate resolutions ($\sim$42 pc), while N2 and O1 remain relatively stable, with smaller and less significant deviations at both $\sim$42 pc and $\sim$170 pc. These results indicate that the ionization conditions in the SMC are strongly structured on scales below $\sim$50 pc, becoming progressively homogenized at scales of $\sim$200 pc, with EW(H$\alpha$) and gas velocity dispersion ($\sigma_{H\alpha}$) remaining the most sensitive tracers of spatial resolution.

In contrast, in the LMC, sensitivity to spatial degradation is more pronounced only at the coarsest resolution ($\sim$272 pc), where S2 and N2 exhibit the largest deviations, whereas O3, O1, EW(H$\alpha$), and $\sigma_{H\alpha}$ are largely robust. Overall, these results highlight that low-ionization line ratios (S2, N2) are more strongly affected by spatial resolution than high-ionization tracers (O3, O1) or kinematic and equivalent width parameters. This suggests that the statistical properties of the ionized ISM in the LMC are largely preserved to scales of $\sim$70 pc, with significant homogenization occurring only above $\sim$250 pc, while in the SMC small-scale structure dominates below $\sim$50 pc.

Overall, the 2D KS results show that the joint distributions of emission-line diagnostics in the SMC are highly sensitive to spatial resolution. Significant changes are already evident at intermediate scales ($\sim$42 pc), indicating that correlations between high- and low-ionization tracers, as well as between ionization, stellar, and kinematic properties, are strongly structured on small scales. At coarser resolution ($\sim$170 pc), several diagrams particularly O3 vs S2, O3 vs O1, WHAN, and WHaD remain significantly affected, reflecting the mixing of distinct ionization regimes, while abundance-related tracers (O3 vs N2 and s2 vs n2) are more resilient to spatial averaging.

For the LMC, most diagnostic diagrams are largely unaffected at intermediate resolution ($\sim$68 pc), except for s2 vs n2, which shows a significant deviation even at this scale. At the coarsest resolution ($\sim$272 pc), all diagrams except WHaD show strong and systematic deviations, with the largest KS values found in O3 vs S2 (KS = 0.1605), s2 vs n2 (KS = 0.1531), and WHAN (KS = 0.1494). By contrast, WHaD remains relatively stable (p = 0.1000; KS = 0.0733), confirming its lower sensitivity to spatial resolution. These results demonstrate that ionization and density sensitive diagnostics are most affected by coarse spatial averaging, whereas parameters related to kinematics ($\sigma_{H\alpha}$) and equivalent width (EW(H$\alpha$)) are more robust across resolutions.

In summary, both the 1D and 2D KS tests show that diagnostic diagrams involving emission-line ratios (O3, S2, N2, O1) are the most sensitive to spatial resolution. In the SMC, significant deviations already appear at $\sim$42 pc, with strong differences persisting at $\sim$170 pc, especially in O3 vs S2, O3 vs O1, WHAN, and WHaD. In the LMC, the distributions remain largely consistent at $\sim$68 pc, while stronger deviations emerge at $\sim$272 pc, particularly for O3 vs S2, s2 vs n2, and WHAN. In contrast, WHaD shows the smallest deviation among the LMC diagnostics at the coarsest resolution (KS=0.0733, p=0.1000), indicating a comparatively lower sensitivity to spatial averaging in the LMC. These results highlight that the impact of spatial resolution is not uniform across diagnostics or galaxies, with the SMC generally showing stronger sensitivity to spatial degradation than the LMC.

\section{Discussion}

\subsection{Interpretation of diagrams as a function of scale}

Our results show several points of convergence with the trends reported by the CALIFA study regarding the impact of spatial resolution on diagnostic diagram behaviour, but also highlight important differences that arise when analysing different components in the Magellanic Clouds (ionized regions and diffuse component). In the work by \citet{mastrosalessanchez2014}, two levels of spatial degradation were applied to five CALIFA galaxies, with physical scales ranging from $\sim$90 pc to 1 kpc. Although their analysis focuses mainly on the classical BPT diagram (and only as a function of galactocentric distance) the general behaviour they report is consistent with what we find across six diagnostic diagrams. In both studies, the highest spatial resolution yields the broadest coverage of the diagnostic parameter space, whereas coarser resolutions collapse the distribution toward the mean values of the original high-resolution data. This behaviour reflects the luminosity-weighted mixing effect described in their work, whereby spatial averaging suppresses the extremes of the distribution and drives the measurements toward intermediate ionization conditions.

However, important differences also emerge. In the CALIFA galaxies analyzed by \cite{mastrosalessanchez2014}, moderate levels of spatial degradation still preserve much of the overall structure of the BPT diagram, but stronger degradation causes the data to contract substantially into the densest part of the distribution. In contrast, the LMC and SMC exhibit more uniform responses across different diagnostic diagrams and remain largely confined to the star-forming ionization regime even at coarse resolutions. We attribute this behavior not only to the significant difference in spatial resolving power between the LVM and CALIFA datasets, but also to the range of physical scales probed: while our analysis reaches spatial scales of only $\sim$300 pc, the CALIFA study explored degradations up to kpc scales, where spatial averaging mixes fundamentally different galactic environments. Additionally, whereas the CALIFA sample shows a clear disappearance of points occupying the so-called composite region at low resolution, our low-metallicity systems retain extensions toward harder-ionization regimes in some diagrams (e.g. O3–S2 and O3–O1), suggesting that the robustness of diagnostic structures to resolution loss depends sensitively on both intrinsic ISM conditions and the physical scale over which emission is averaged. Thus, although both studies demonstrate that spatial averaging reduces the range of values spanned by diagnostic diagrams, the Magellanic Clouds show comparatively more stable ionization signatures across scales, indicating a weaker resolution dependence in low-mass, metal-poor environments within the sub-kpc regime explored here.

Our diagnostic diagrams show that the main effect of degrading spatial resolution is not a wholesale displacement of the central star-forming locus, but a loss of the extreme line-ratio and kinematic tails, especially outside bright H$\alpha$ structures. This behavior is consistent with resolved IFS studies showing that DIG systematically enhances low-ionization ratios such as [NII]/H$\alpha$, [SII]/H$\alpha$, and [OI]/H$\alpha$, shifting spaxels toward composite or LINER-like regions, while classical HII region line ratios remain comparatively stable under spatial rebinning \citep{zhangyanbundy2017, poetrodjojodagostinogroves2019}. The low-EW(H$\alpha$), broader-velocity-dispersion DIG sequence seen in our external DIG pixels is also in line with resolved studies that associate DIG with harder ionization conditions and, in some systems, elevated turbulence or additional non-photoionization contributions \citep{belfioresantoerogroves2022}. The systematically lower N2, S2, and O1 values in the SMC relative to the LMC are moreover qualitatively consistent with the lower metallicity of the SMC, while the broader parameter range in the LMC agrees with evidence that the LMC contains substantial spatial variation in nebular conditions and gas kinematics \citep{lahcollessdeugenio2024}. Overall, the comparison suggests that the highest-resolution maps recover a more heterogeneous mixture of HII region and DIG conditions, whereas coarser resolution increasingly homogenizes the distributions through luminosity-weighted mixing and blurred HII/DIG boundaries \citep{poetrodjojodagostinogroves2019}.

While the general effect of spatial degradation can be understood in terms of the increasing mixing between HII regions and DIG, the extended high $\sigma_{\mathrm{H}\alpha}$ tail observed in the SMC WHaD diagram may have an additional physical origin. In particular, the presence of rapidly rotating Be stars provides a possible explanation for the high velocity dispersions observed in this tail.

In \cite{martayanfloquethubert2007}, a catalogue of Be-type stars in the SMC is presented. The authors investigate how the high rotational velocities characteristic of this class of stars affect the emission profile of the H$\alpha$ line, producing significant line broadening. They also compare the occurrence of Be stars in different environments, including the LMC and the Milky Way.

Their analysis shows that the fraction of field Be stars relative to B-type stars is slightly higher in the SMC ($\sim$26\%) than in the LMC ($\sim$17.5\%) and in the Milky Way ($\sim$15\% for single-epoch surveys). Consequently, the decrease in metallicity appears to play a significant role in enhancing the occurrence of the Be phenomenon.

We suggest that the presence of Be stars may be the physical mechanism responsible for the tail of the contour distribution observed in the SMC case in the WHaD diagnostic diagram, where the velocity dispersion values can reach up to $\sim$100 km s$^{-1}$.

\subsection{Separation of HII regions and DIG}

We use the \texttt{pyHIIextractor} code to identify clumpy ionized regions and separate the DIG component, defined as pixels outside the detected clumpy regions. Although the code provides a model for the DIG, a more detailed analysis of this component will be addressed in future work, which will also allow the application of additional methods for its characterization and modeling.

In contrast to the validation of \texttt{pyHIIextractor} performed by \cite{lugosanchezespinosa2022} using simulations of spiral galaxies our application of the code targets a very different physical regime: the Magellanic Clouds as observed with LVM, where the achieved spatial scales allow us to directly resolve the structure of the ionized gas. In the spiral-galaxy simulations, the recovered line ratios exhibit systematic biases of a few percent in N2 and O3, and larger discrepancies in EW(H$\alpha$), primarily driven by projection effects and the overlap of structurally distinct regions because the explored resolution ranges are much lower ($>$300 pc). In our case, the proximity of the Magellanic Clouds enables a more stable recovery of line ratios dominated by photoionization from young stars. Although our identification of ionized regions relies on circular apertures, this approach effectively captures the global behavior of the line ratios, as reflected in the six diagnostic diagrams explored in this work. Nevertheless, achieving a more detailed characterization of the internal structure of the regions particularly for studies sensitive to leakage, geometry, or spatial variations will ultimately require methods capable of recovering more complex morphologies than those imposed by circular apertures.

The behavior of the Magellanic Clouds across our diagnostic diagrams offers a contrast to the DIG framework proposed by \cite{lacerdafernandescouto2018}. Their CALIFA-based study, conducted at kiloparsec scales, showed that EW(H$\alpha$) is an effective discriminator of ionization regimes and that hDIG (provided by Hot Low-Mass Evolved Stars or HOLMES, \citealp{floresfajardomorissetstasinska2011}: EW(H$\alpha$) $<$ 3 \AA), mDIG (mixed ionization: 3 \AA $<$ EW(H$\alpha$) $<$ 14 \AA), and SFc (star-forming complexes: EW(H$\alpha$) $>$ 14 \AA) form a continuous sequence driven by the relative contributions of old stars and young stars. In our case, the much smaller physical scales probed by LVM (less than $\sim$40 pc) imply that the ionization is overwhelmingly dominated by young stellar populations, even in regions outside the apertures identified as ionized regions. As a result, EW($\alpha$) and $\sigma_{H\alpha}$ remain remarkably stable with resolution, and the diffuse emission tends to stay closer to the star-forming locus rather than shifting toward HOLMES dominated regimes. On the other hand, we cannot rule out the possibility that, due to the S/N cut adopted in this work, the most DIG associated with ionization by HOLMES is systematically excluded from our analysis. This effect may be further amplified when considering equivalent-width-based diagnostics. Since the equivalent width is defined as the ratio between the emission-line flux and the underlying stellar continuum, both quantities can be affected by the spatial resolution. In particular, while the line emission may arise locally within a $\sim$50 pc spaxel, the stellar continuum may originate from a more spatially extended stellar population, potentially biasing the measured EW values. However, this remains a hypothesis to be explored in future work, in which different methodologies for the characterization of the DIG will be investigated.

The PHANGS sample analyzed by \cite{belfioresantoerogroves2022} yields a significantly lower mean EW($\alpha$) of $\sim$ 5 \AA\ for the diffuse component due to the leakage of ionizing photons from HII regions. In contrast, the mean EW($\alpha$) values traced by our density contours align more closely with the high-EW($\alpha$) diffuse component described by \cite{lacerdafernandescouto2018} rather than with the lower-EW($\alpha$) DIG regime characteristic of PHANGS galaxies. Although the PHANGS–MUSE observations are optimized for high spatial resolution, they reach a comparatively shallower surface-brightness sensitivity than the Local Volume Mapper data. Since EW(H$\alpha$) depends on the fraction of diffuse ionized emission recovered, both sensitivity differences and the adoption of an S/N(H$\alpha$) $>$ 3 threshold in our analysis preferentially select the brighter component of the diffuse gas while excluding the lowest surface-brightness emission. This selection effect naturally shifts the measured EW(H$\alpha$) distribution toward higher values.


All together, these comparisons highlight that the interpretation of diffuse ionization depends sensitively on both the intrinsic properties of galaxies and the spatial resolution of the observations, and may also depend on the adopted signal-to-noise threshold.

In general, the diagnostic diagrams corresponding to all the pixels of the LMC and SMC mostly fall within the star-forming region in all six diagrams. This behavior is expected, given that both galaxies exhibit high star-formation activity. Nevertheless, in the “composite’’ regions of the O3 vs. N2, O3 vs. S2, and O3 vs. O1 diagrams (as discussed by \cite{kewleyellison2008}), the presence of values does not necessarily indicate an ionization source beyond star formation. Other contributors, such as shocks or HOLMES (e.g., post-AGB stars), acting together with star formation, can also populate these regions \citep{floresfajardomorissetstasinska2011}.

There are many physical processes occurring simultaneously that can produce very similar emission-line ratios, as noted by \cite{sanchez2020}, and the location of points in diagnostic diagrams is further shaped by global properties such as metallicity. This is particularly relevant for the SMC and LMC, whose low metallicities naturally influence their placement in diagrams like O3 vs N2. In this context, the work of \cite{bresolinkennicuttryan2012}, which presents the \cite{dopitafischera2006} models for the O3 vs N2 diagram, is especially informative: although focused on HII regions, it shows that values with N2 $<$ –1 typically correspond to subsolar metallicities (consistent with the SMC and, to a lesser extent, the LMC). While we do not quantify these effects in the present study, they underscore the importance of metallicity- and age-dependent ionization conditions. A more exhaustive treatment will require ionization modeling across all diagnostic diagrams, which we plan to pursue in future work.

Several studies have investigated the emission-line properties of low-metallicity galaxies at high redshift. For example, the work of \cite{zavalacastellanoakins2025} presents the classical BPT diagnostic diagram and compares the observations with stellar photoionization models from \cite{nakajimamaiolino2022}; (see their Figure 2).

If we adopt the central values of our emission-line distributions in the classical BPT diagram (blue contours corresponding to all pixels in Figures \ref{fig:smc_diagrams} and \ref{fig:lmc_diagrams} of this work) as representative average values for the SMC and the LMC, we find qualitative agreement with the mean values reported by \cite{zavalacastellanoakins2025} for systems with metallicities of approximately 0.2 Z$_\odot$ (SMC) and 0.5 Z$_\odot$ (LMC). This comparison suggests an ionization parameter of approximately log U $\sim$ -3 for both galaxies, inferred by comparing the observed line ratios with predictions from photoionization models powered by star formation \citep{dagostinokewleygroves2019, toppingshappleyreddy2020}. Such an average ionization level is primarily associated with photoionization driven by ongoing star formation, placing the contour distributions below and to the right of the standard demarcation curves in the BPT diagram. This result also highlights the role of key physical parameters, particularly metallicity and ionization parameter, in shaping the location of galaxies within diagnostic diagrams. A more detailed exploration of these dependencies will require comprehensive future modeling efforts.

In Fig. \ref{fig:smc_diagrams} and \ref{fig:lmc_diagrams}, we show that, in general, the clumpy ionized regions are associated with star-forming regions. However, our selection of clumpy ionized regions is based on the H$\alpha$ line, and not only young and hot stars produce significant H$\alpha$ emission. Therefore, while we suggest that star formation dominates in the LMC and SMC, a more exhaustive study of all types of ionizing sources is required, including a characterization by stellar population types, morphology, chemical properties, and their dynamics.

Emission-line ratios such as [NII]$\lambda$6583/H$\alpha$, [OIII]$\lambda$5007/H$\beta$ and [SII]$\lambda$$\lambda$6717+6731/H$\alpha$ are widely used to probe the ionization conditions and gas-phase oxygen abundance, but their diagnostic power depends critically on spatial resolution. As resolution degrades, the mixing of regions with intrinsically different ionization conditions washes out the fine scale structure present in the highest resolution data, biasing line ratios toward intermediate values. This loss of contrast can directly affect metallicity estimates derived from strong-line methods, especially given the systematic differences among commonly used calibrations (e.g. \citealp{pettinipagel2004}; \citealp{marinorosalessanchez2013}; \citealp{pilyugingrebel2016}; \citealp{valeasaricoutocidfernandes2019}). These issues highlight the importance of characterizing how line ratio diagnostics behave at the native, finest available resolution before interpreting measurements obtained at coarser scales, where physical information may be partially or entirely blended.

\section{Conclusions}

Our results demonstrate that spatial resolution degradation impacts diagnostic diagrams in different ways. High resolution BPT diagrams retain a larger fraction of the physical information associated with distinct ionization conditions, whereas WHAN and WHaD show a more complex, galaxy-dependent response to spatial degradation. At coarse spatial resolution, luminosity-weighted mixing progressively reduces the contribution of localized extreme conditions and drives the distributions toward more intermediate values. However, the degree of homogenization differs between the SMC and LMC, with the SMC showing a stronger response to spatial degradation. At spatial scales below $\sim$50 pc, the BPT diagrams resolve a wider range of ionization conditions that can be studied individually. Taken together, these results emphasize that high spatial resolution is essential to recover the full diversity of ionized-gas physical conditions and that the impact of spatial degradation cannot be characterized by a single response applicable to all diagnostics or galaxies.

\begin{itemize}

    \item As spatial resolution decreases, the number of detected ionized regions drops sharply, with an average loss of $\sim$82\% of detections at each degradation step, while the mean integrated H$\alpha$ flux of the remaining regions increases systematically by factors of $\sim$7-11 at intermediate scales and $\sim$3 at coarser scales, reflecting the progressive blending and disappearance of compact structures.

    \item Cross-resolution comparisons show that spatial degradation leads to substantial merging of compact regions: in both the SMC and LMC, individual low-resolution detections contain up to $\sim$12-18 high-resolution regions at the coarsest scales. At the same time, the fraction of high-resolution regions recovered within low-resolution detections decreases, indicating strong structural mixing and the progressive loss of compact regions into diffuse emission.

    \item The integrated H$\alpha$ flux attributed to the DIG increases dramatically with spatial degradation, reaching enhancements of more than an order of magnitude at the coarsest scales, demonstrating the strong sensitivity of DIG measurements to resolution effects.
    
    \item Emission-line ratios (O3, S2, N2, O1) are the parameters most affected by spatial resolution, as evidenced by the largest deviations in both the 1D and 2D KS tests.

    \item The one- and two-dimensional KS tests show that the sensitivity to spatial resolution depends on both the diagnostic parameter and the galaxy. In the SMC, O3, EW(H$\alpha$), and S2 show the strongest differences at the intermediate resolution of $\sim$42 pc, while EW(H$\alpha$) remains significantly affected at $\sim$170 pc. In the LMC, the largest deviations occur at the coarsest resolution of $\sim$272 pc, particularly for S2 and N2, whereas O3, EW(H$\alpha$), and $\sigma_{\mathrm{H}\alpha}$ show comparatively smaller differences. Thus, no individual parameter can be considered universally robust or universally sensitive to spatial degradation. The 1D tests quantify changes in the distributions of individual parameters, while the 2D tests reveal how spatial averaging modifies the correlations between the quantities defining the diagnostic diagrams.
    
    \item SMC is highly structured on small scales and intrinsically more susceptible to resolution-driven blending, with significant differences emerging already at intermediate resolutions ($\sim$42–170 pc), whereas in the LMC the ionized medium is more spatially coherent, showing noticeable effects only at the coarsest scale ($\sim$272 pc).
    
    \item Decreasing spatial resolution produces a homogenization effect: fine-scale variations and extreme values present at the highest resolutions (17-19 pc) are progressively diluted, causing lower-resolution distributions to collapse toward the mean.
    
    \item The DIG is the component most strongly affected by spatial degradation; its filamentary and low-surface-brightness structures blend with neighbouring regions, artificially shifting diagnostic diagrams toward harder ionization regimes.
        
    \item Highest resolutions (below $\sim$40 pc) capture the full diversity of physical conditions present in both the ionized regions and the diffuse gas, preserving extreme values, sharp spatial gradients, and localized structures that are otherwise lost. At intermediate (42–68 pc) and coarse (170–272 pc) scales, these features are progressively smoothed out, and the resulting distributions sample only restricted subsets of the parameter space, missing the tails and reducing the apparent dynamic range of key diagnostics.
    
    \item The response of the WHaD diagram to spatial degradation is galaxy-dependent. In the SMC, the WHaD distribution shows statistically significant differences at both $\sim$42 and $\sim$170 pc, indicating that the combined distribution of EW(H$\alpha$) and $\sigma_{\mathrm{H}\alpha}$ is sensitive to spatial averaging in this galaxy. In contrast, the LMC shows a comparatively stable WHaD distribution, with the smallest deviation among the 2D diagnostic diagrams at the coarsest resolution ($\sim$272 pc; KS = 0.0733, $p=0.1000$). Therefore, the relative robustness of WHaD cannot be generalized to both galaxies, but depends on the intrinsic structure of the ionized gas and on the spatial scale considered.
    
\end{itemize}

High spatial resolution ($<$50pc) is essential to capture the full diversity of the ionized ISM, as coarser resolutions systematically average over structures, erasing fine details and affecting both emission-line diagnostics and the characterization of clumpy ionized regions and the DIG.

\section*{Acknowledgements}

AZL-A gratefully acknowledges the support provided by the Postdoctoral Program (POSDOC) of UNAM (Universidad Nacional Autónoma de México). HIM acknowledge the support from grants SECIHTI CBF2023-2024-1418 and CF-2023-G-543, PAPIIT UNAM IA104325, IN-106823, and IN-119123. JEM-D and AZL-A thanks the support by SECIHTI CBF-2025-I-2048 project “Resolviendo la Física Interna de las Galaxias: De las Escalas Locales a la Estructura Global con el SDSS-V Local Volume Mapper”. CR-Z acknowledges support from project UNAM DGAPA-PAPIIT IG 101723. R de JZ would like to express his appreciation for the support provided by SECIHTI (Secretaría de Ciencia, Humanidades, Tecnología e Innovación) in form of their Posdoctoral Grant. MAV-D was supported by by SECIHTI (Secretaría de Ciencia, Humanidades, Tecnología e Innovación) in form of Beca Nacional de Posgrado. GAB acknowledges the support from the ANID Basal project FB210003. IC-G acknowledges support from the UNAM-DGAPA-PAPIIT grants IN-119123 and IN-113726, and from SECIHTI grant CF-2023-G-100. PJH acknowledges support from Universidad Nacional Aut{\'o}noma de M{\'e}xico Postdoctoral Program (POSDOC). KK, EE gratefully acknowledge funding from the Deutsche Forschungsgemeinschaft (DFG, German Research Foundation) in the form of an Emmy Noether Research Group (grant number KR4598/2-1, PI Kreckel) and the European Research Council’s starting grant ERC StG-101077573 (“ISM-METALS"). CM acknowledges financial support from the UNAM DGAPA grant IG101233. TH is supported by an Australian Government Research Training Program (RTP) Scholarship. JGF-T gratefully acknowledges the support provided by ANID Fondecyt Regular No. 1260371. OE acknowledges funding from the Deutsche Forschungsgemeinschaft (DFG, German Research Foundation) -- project-ID 541068876.

Funding for the Sloan Digital Sky Survey V has been provided by the Alfred P. Sloan Foundation, the Heising-Simons Foundation, the National Science Foundation, and the Participating Institutions. SDSS acknowledges support and resources from the Center for High-Performance Computing at the University of Utah. SDSS telescopes are located at Apache Point Observatory, funded by the Astrophysical Research Consortium and operated by New Mexico State University, and at Las Campanas Observatory, operated by the Carnegie Institution for Science. The SDSS web site is \url{www.sdss.org}. SDSS is managed by the Astrophysical Research Consortium for the Participating Institutions of the SDSS Collaboration, including Caltech, The Carnegie Institution for Science, Chilean National Time Allocation Committee (CNTAC) ratified researchers, The Flatiron Institute, the Gotham Participation Group, Harvard University, Heidelberg University, The Johns Hopkins University, L’Ecole polytechnique f\'{e}d\'{e}rale de Lausanne (EPFL), Leibniz-Institut f\"{u}r Astrophysik Potsdam (AIP), Max-Planck-Institut f\"{u}r Astronomie (MPIA Heidelberg), Max-Planck-Institut f\"{u}r Extraterrestrische Physik (MPE), Nanjing University, National Astronomical Observatories of China (NAOC), New Mexico State University, The Ohio State University, Pennsylvania State University, Smithsonian Astrophysical Observatory, Space Telescope Science Institute (STScI), the Stellar Astrophysics Participation Group, Universidad Nacional Aut\'{o}noma de M\'{e}xico, University of Arizona, University of Colorado Boulder, University of Illinois at Urbana-Champaign, University of Toronto, University of Utah, University of Virginia, Yale University, and Yunnan University.

\section*{Data Availability}
\label{sec:data_ava}

The LVM data used in this work will be made publicly available with SDSS Data Release 21. Custom scripts developed for cube reconstruction and the \texttt{pyHIIextractor} analysis are available from the authors upon reasonable request.

\bibliographystyle{mnras}
\bibliography{example}

\bsp	
\label{lastpage}
\end{document}